\documentclass[letterpaper,11pt]{article}
\usepackage{jheppub}

\usepackage[utf8]{inputenc}
\usepackage{feynmp-auto}
\usepackage{graphicx}
\usepackage{graphbox}
\usepackage{amsmath}
\usepackage{amsfonts}
\usepackage{amssymb}
\usepackage{amsthm}
\usepackage{mathtools}
\usepackage{physics}
\usepackage[dvipsnames]{xcolor}
\usepackage{float}
\usepackage{caption}
\usepackage{subcaption}
\usepackage{comment}
\usepackage{mathrsfs}
\usepackage{tensor}
\usepackage[export]{adjustbox}
\usepackage{tikz}
\usepackage{latexsym}
\usepackage{slashed}
\usepackage{orcidlink}
\usetikzlibrary{hobby,calc,shapes, positioning, backgrounds,trees, decorations, decorations.markings, decorations.pathmorphing, arrows.meta, shapes.geometric, snakes}

\def\be{\begin{equation}}
\def\ee{\end{equation}}
\def\ba{\begin{aligned}}
\def\ea{\end{aligned}}

\def\p{\partial}

\def\oh{\mathcal{O}}
\def\ah{\mathcal{A}}
\def\nh{\mathcal{N}}
\def\uh{\mathcal{U}}
\def\fh{\mathcal{F}}
\def\zh{\mathcal{Z}}
\def\kh{\mathcal{K}}

\def\bi{\begin{itemize}}
\def\ei{\end{itemize}}
\def\it{\textit}

\def\mbf{\boldsymbol}

\def\e_1{\epsilon_1}
\def\e_2{\epsilon_2}
\def\e{\epsilon}
\def\Trop{\text{Trop}}
\DeclareMathAlphabet{\mathbbold}{U}{bbold}{m}{n}
\newcommand{\nn}{\nonumber}
\renewcommand{\d}{\mathrm{d}}

\newcommand{\PF}{{\mathcal{P} \! F}}
\newcommand{\Ptr}{{\mathcal{P} \! \tr}}
\newcommand{\Pgam}{{\mathcal{P} \! \Gamma}}
\newcommand{\Trsu}{\mathrm{Tr}}

\title{Surfaceology for Colored Yukawa Theory}

\author[a, \orcidlink{0000-0002-2013-3201}]{Shounak De,}
\emailAdd{shounak\_de@brown.edu}

\author[a, \orcidlink{0000-0003-1186-4624}]{Andrzej Pokraka,}
\emailAdd{andrzej\_pokraka@brown.edu}

\author[a, \orcidlink{0000-0003-2569-1234}]{Marcos Skowronek,}
\emailAdd{marcos\_skowronek\_santos@brown.edu}

\author[a,b,c, \orcidlink{0009-0005-6084-2466}]{Marcus Spradlin,}
\emailAdd{marcus\_spradlin@brown.edu}

\author[a,b, \orcidlink{0009-0008-2506-3207}]{Anastasia Volovich}
\emailAdd{anastasia\_volovich@brown.edu}

\affiliation[a]{Department of Physics, 	
    Brown University, 	
    Providence, 	
    RI 02912, 
    USA
}

\affiliation[b]{
    Department of Physics,
    Harvard University,
    Cambridge, 
    MA 02138, 
    USA
}

\affiliation[c]{Brown Theoretical Physics Center,
    Brown University,
    Providence,
    RI 02912,
    USA
}

\abstract{Arkani-Hamed and collaborators have recently shown that scattering amplitudes for colored theories can be expressed as integrals over combinatorial objects simply constructed from surfaces decorated by kinematic data. In this paper we extend the curve integral formalism to theories with colored fermionic matter and present a compact formula for the all-loop, all-genus, all-multiplicity amplitude integrand of a colored Yukawa theory. The curve integral formalism makes certain properties of the amplitudes manifest and repackages non-trivial numerators into a single combinatorial object. We also present an efficient formula for $L$-loop integrated amplitudes in terms of a sum over $2^L$ combinatorial determinants.}

\begin{document}\maketitle

\section{Introduction}

Recently Arkani-Hamed and collaborators \cite{Arkani-Hamed:2023lbd, Arkani-Hamed:2023mvg, Arkani-Hamed:2024vna} have reformulated scattering amplitudes for colored theories in terms of fundamentally geometric/combinatorial objects called curve integrals. 
Remarkably, in a theory of matrix-valued scalars, the color-ordered partial amplitude at any loop order and multiplicity is computed via a global Schwinger integral that is the tropicalization of certain \emph{stringy} integrals. 
These objects provide a unified description for all Feynman diagrams of any given topology, as well as a canonical definition of loop integrands. The curve integral perspective has facilitated a host of novel results: 
hidden zeroes \cite{Bartsch:2024amu, Arkani-Hamed:2023swr}, 
new factorization relations \cite{Arkani-Hamed:2024fyd, Cao:2024gln}, gluons from scalars \cite{Arkani-Hamed:2023jry}, the relationship between $\Trsu(\Phi^3)$ and non-linear sigma model amplitudes \cite{Arkani-Hamed:2024nhp, Arkani-Hamed:2024yvu}, 
and related work \cite{Laddha:2024qtn,GimenezUmbert:2024jjn,Dong:2024klq,Li:2024qfp}.

In this paper, we extend the curve integral formalism to theories with fermionic matter; an essential step towards realizing the amplitudes of more general quantum field theories in terms of purely combinatorial/geometric building blocks. 
Specifically, we study a colored Yukawa theory and derive compact tropical expressions for  $L$-loop $n$-point amplitudes both before and after loop integration. 
Our formulae do not make reference to a sum over Feynman diagrams and only depend on the combinatorics of curves associated to distinguished ribbon graphs or fatgraphs.
In particular, our post loop-integration formula is a $2^L$-term sum over tropical determinants that encode the tensor structure of the theory.
Despite avoiding the need to sum over a large number of Feynman diagrams, our formulae have an inherent $2^L$ growth in complexity stemming from the fact that each internal puncture in the surface can be assigned different species (in this case, two). 

The paper is organized as follows. Section \ref{sec:review} reviews the essential aspects of the curve integral formalism, highlighting the essential steps to compute the relevant geometric objects. 
Section \ref{sec:Yukawa} defines the Lagrangian for our colored Yukawa theory and 
explains how the color ordering of the interaction vertex places constraints on the theory. 
In subsection \ref{subsec:combinatorialamp}, we present the main result of this work: a curve integral formula for colored Yukawa amplitudes.
Section \ref{sec:examplitudes} provides pedagogical examples at tree and loop level demonstrating the practical application of our formulae. 
Section \ref{sec:Outlook} concludes with potential directions for future research.


\section{Surfaceology review}
\label{sec:review}

In this section, we review the basic aspects of the curve integral formalism, outlining the necessary steps to compute the ``curvy'' objects that enter our formulae for scattering amplitudes in colored Yukawa theory. 
For more insight into the physical and mathematical significance of these elements, we refer to the original works and their extensions \cite{Arkani-Hamed:2023lbd, Arkani-Hamed:2023mvg, Arkani-Hamed:2024vna, Arkani-Hamed:2023swr, Arkani-Hamed:2023jry, Arkani-Hamed:2024nhp, Arkani-Hamed:2024yvu,Arkani-Hamed:2024fyd}.

\begin{figure}[h]
    \centering
    \includegraphics[width=14cm,valign=c]{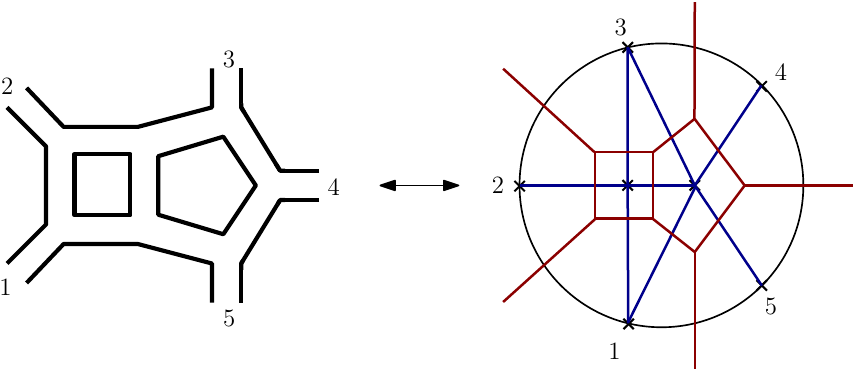}
    \caption{Surface triangulation (right) corresponding to a certain fatgraph (left). Both are related by graph duality.}
    \label{fig:graph duality} 
\end{figure}

In the curve integral formalism, the kinematic data of color-ordered partial amplitudes is described by the set of all possible curves on a representative fatgraph $\Gamma$ for the amplitude in question. 
It is also often convenient to work with the associated surface $\Sigma$ where the fatgraph $\Gamma$ corresponds to a specific triangulation of $\Sigma$ via graph duality (see figure \ref{fig:graph duality}).
Given a fatgraph $\Gamma$, its propagators can be interpreted as roads along which certain paths or \emph{curves} are drawn. 
Each road is assigned a variable $y_e$, and a curve $C$ is described by a \emph{word} which records the series of roads and left/right turns at a vertex that the curve $C$ makes on the fatgraph
\be C = y_iRy_jLy_k\ldots.\ee
On the surface, the endpoints of a path can either be an external or an internal puncture. In the latter case, this is depicted on the fatgraph as spiraling indefinitely around a loop. 
Due to an over-counting associated with the direction of the spiral, we choose to work exclusively with curves that spiral \emph{counterclockwise} (if one chooses to work with the clockwise convention, some formulae must be modified).

From these curves, we generate the kinematic data of color-ordered partial amplitudes.
The momentum associated with each curve $P^\mu_C$ is given by the prescription
\be P_C^\mu = P^\mu_\text{start} + \sum_\text{right turns} P^\mu_\text{incoming from the left}. \label{eq:momlabel} \ee
In other words, each time the path takes a right turn, we add the incoming momentum from the left road of the vertex.
We also introduce the $g$-vector associated to each curve, $\vec{g}_C\in\mathbb{R}^E$, where $E$ is the number of roads in the fatgraph. 
In order to compute it, it is useful to represent the left/right turns in the path as going up/down along a \textit{mountainscape}
\be C = y_i R y_j L y_k L y_l R y_m\ldots \quad \longleftrightarrow \quad
    \includegraphics[width=6cm,valign=c]{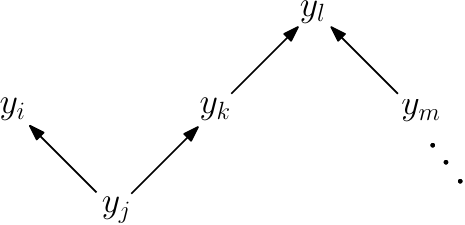}
\ee
Then, the components of the corresponding $g$-vector are
\be g_C^e = (\#\, \text{of peaks at}\, e) - (\#\, \text{of valleys at}\, e).\ee
The collection of $g$-vectors divides $\mathbb{R}^E$ into a spanning set of top-dimensional cones that defines the \emph{Feynman fan}. 
Since each top-dimensional cone of the Feynman fan corresponds to a specific Feynman diagram, the propagators of a Feynman diagram are those associated to the $g$-vectors defining that cone.

In order to determine which curves are present in the different cones on the fan, we need another object called the \emph{tropical headlight function} $\alpha_C(\mbf{t})$. 
These can be computed from the associated $2\times2$ curve matrices $M_C$, which encode the information of each curve $C$. We obtain them by replacing the elements in the words with the associated matrices
\be y_e\to \begin{pmatrix}
    1&0\\0&y_e
\end{pmatrix},\ L\to \begin{pmatrix}
    1&0\\1&1
\end{pmatrix}, \ R\to \begin{pmatrix}
    1&1\\0&1
\end{pmatrix}.\ee
Then, the corresponding headlight function $\alpha_C(\mbf{t})$ is given by
\be \alpha_C(\mbf{t}) = \Trop [(M_C)_{11}] + \Trop [(M_C)_{22}] 
	- \Trop [(M_C)_{12}] - \Trop [(M_C)_{21}],\ee 
where the tropicalization of a function $f(\mbf{y})$ is defined as the limit
\be f(\mbf{y})\to \exp\big[\Trop(f)(\mbf{t})\big],\quad \text{as}\quad y_e = \exp t_e,\ t_e\to\infty.\ee
For computational purposes, this simply amounts to the replacement rule
\be x y\to t_x + t_y,\quad x+y\to \max(t_x,t_y).\ee 
Importantly, the headlight functions $\alpha_C$ are piecewise linear in each cone on the Feynman fan. Moreover, they only ``light up'' on the cones bounded by the corresponding vector $\vec{g}_C$
\be \alpha_C(\vec{g}_D) = \delta_{C,D}.\ee
Another useful function that is directly related to the $\alpha_C$'s is the \emph{tropical step function} $\Theta_C(\mbf{t})$, which is piecewise constant on each cone  \cite{Arkani-Hamed:2024vna}
\be \Theta_C 
= \vec{g}_C\cdot \nabla\alpha_C
= \begin{cases}
    1& \text{in any cone that contains $C$},\\
    0&\text{otherwise}.
\end{cases}\ee

Starting from non-planar surfaces at one loop (or at two loops for planar ones), there are sets of curves that can wind an arbitrary number of times around the punctures before ending. 
These curves correspond to the same Feynman diagrams and therefore, it is necessary to mod out by the Mapping Class Group (MCG) so that we do not over count \cite{Arkani-Hamed:2023lbd}. 
This is achieved at the level of the curve integral by inserting a tropical function $\mathcal{K}$ called the \emph{Mirzakhani kernel}. In the examples considered in this paper, there is only one generator of the MCG, which takes the form of a closed curve $\Delta$\footnote{For a more general treatment of this process, please refer to \cite{Arkani-Hamed:2023lbd}.}.  The curves that intersect $\Delta$ are divided into different cosets under the MCG. For each coset $i$, we choose a curve representative $C_i^0$ and keep only the paths compatible with this set. 
Using this data, the Mirzakhani kernel is 
\be \kh = \sum_i \frac{\alpha_{C_i^0}}{\rho},\ee
where $\rho$ is the sum over the set $\mathcal{S}$ of all curves not invariant under MCG
\be \rho = \sum_{C\in\mathcal{S}}\alpha_C.\ee
Since $\kh$ has support on a finite set of cones, we only need to compute a small number of headlight functions, namely the ones compatible with the coset representatives $C_i^0$.

\section{Fermionic curve integrals for colored Yukawa theory}
\label{sec:Yukawa}

This section starts with the Lagrangian for our colored Yukawa model and enumerates the basic Feynman rules. 
Subsequently, we explain how the color-ordered partial amplitudes are subject to stringent constraints that are made manifest in the surface picture. 
These constraints greatly simplify the structure of the partial amplitudes. 
Lastly, we present a formula for the loop integrands and their integrated partial amplitudes in our colored Yukawa theory.

\subsection{Colored Yukawa theory}
\label{sec:YukawaLagrangian}
In this paper, we consider $N^2$ massless colored Dirac fermions and scalars transforming in the fundamental $\times$ anti-fundamental representation of $SU(N)$ minimally coupled via a cubic Yukawa and a cubic scalar vertex. Explicitly, the Lagrangian is
\begin{align} \label{eq:ordLagrangian}
    \mathcal{L} &= \Trsu \Bigg[
        (i \bar{\Psi}^{A a} \slashed{\partial}_{AB} \Psi^{B b})
            T^a T^b
        {-} \frac{1}{2} (\partial_\mu\Phi^a \partial^\mu\Phi^b) T^a T^b
        {+} g\ \bar\Psi^{Aa} \Psi_A^b \Phi^c  T^a T^b T^c 
        {+} \lambda\Phi^a\Phi^b\Phi^cT^a T^b T^c\Bigg]~,
    \nn\\ &= 
    i\Trsu\big(\bar\Psi \slashed{\p}\Psi\big) 
    -\frac{1}{2}\Trsu\big(\p_\mu\Phi\p^\mu\Phi\big) 
    + g\Trsu\big(\bar\Psi\Psi\Phi\big) 
    + \lambda \Trsu\big(\Phi^3\big)~. 
\end{align}
Here, $A,B$ are Dirac indices, $T^a$ are the $SU(N)$ generators, $\Phi^i_j = \Phi^a (T^a)^i_j$, and $\Psi^i_j = \Psi^a (T^a)^i_j$ with $i,j = 1,2,\dots, N$.\footnote{In what follows, we denote the trace over $SU(N)$ color indices $a,b$ as $\Trsu[\dots]$ and the trace over Dirac indices $A,B$ as $\tr[\dots].$}
The Feynman rules for the propagators and vertices are 
\be\label{eq:Feynman rules}
    \Delta_{\Psi\bar{\Psi}}^{ab} = \frac{i\slashed{p}}{p^2}\delta^{ab}~,
    \quad
    \Delta_\Phi^{ab} = \frac{-i}{p^2}\delta^{ab}~,
    \quad
     V_Y^{abc} = ig\ 
    \Trsu(T^aT^bT^c)~, 
    \quad 
    V^{abc}_{\Phi^3} = i\lambda \Trsu(T^aT^bT^c)~,
\ee
and we assume all the particles to be ingoing.
In the following, we consider partial ordered amplitudes and drop the dependence on the color indices $a,b,c$.
We also assume that the Dirac fermions and gamma matrices are in arbitrary number of dimensions $D$.

\begin{figure}
    \centering
    \includegraphics[width=10cm,valign=c]{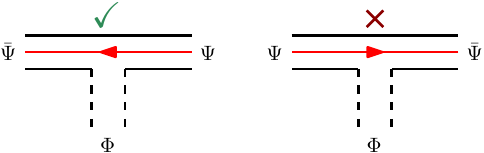}
    \caption{The Lagrangian (\ref{eq:ordLagrangian}) only allows for one ordering of the cubic vertex (left), while the other one (right) cannot appear in the partial amplitudes.}
    \label{fig:allowed vertices} 
\end{figure}

It is very important to note that the Yukawa interaction vertex has a definite color ordering. 
Consequently, the color indices must be contracted in the order dictated by $\Trsu(\bar\Psi\Psi\Phi)$ in double-line diagrams (see figure \ref{fig:allowed vertices}).   
The diagrams that contribute to a partial amplitude are then constrained such that at every interaction vertex, the outgoing fermion has to appear immediately after the outgoing anti-fermion in clockwise order. 
In the language of curves, this means that the fermionic charge flow always turns right at every vertex in the fatgraph.

\begin{figure}
    \centering
    \includegraphics[width=10cm,valign=c]{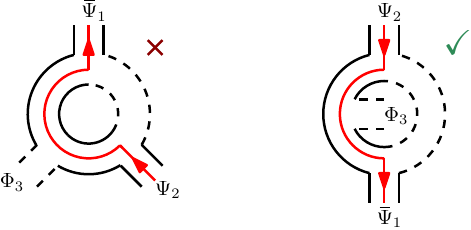} 
    \caption{The planar three-point topology (left) cannot be constructed with the vertex ordering in the Lagrangian (\ref{eq:ordLagrangian}), but the same Feynman integral is obtained from the non-planar diagram (right).}
    \label{fig:non-planar triangle} 
\end{figure}

Compared to a theory where both vertex orderings are allowed, only a subset of diagrams is allowed for each partial amplitude of (\ref{eq:ordLagrangian}). 
However, we want to emphasize that all Feynman integrals that can appear in ordinary (uncolored) Yukawa theory are present in at least one partial amplitude.\footnote{In other words, this formulation (\ref{eq:ordLagrangian}) is consistent with uncolored Yukawa theory, i.e., if we substituted for the $SU(N)$ generators $T^a\to \frac{1}{\sqrt{N}}\mathbbold{1}$, we would recover the amplitudes in conventional Yukawa theory without color indices. Note that one has to keep partial ordered amplitudes that would normally vanish from traceless-ness of the $SU(N)$ generators.}
To illustrate this, consider the one-loop three-point amplitude $\ah^{\text{one-loop}}_3(\bar\Psi_1,\Psi_2,\Phi_3)$. 
It is straightforward to see that it is impossible to draw the fatgraph corresponding to the planar triangle topology involving two fermionic propagators, but the associated Feynman integral can nevertheless be obtained from the double trace, non-planar amplitude $\ah^{\text{one-loop}}_3(\bar\Psi_1, \Psi_2|\Phi_3)$ (see figure \ref{fig:non-planar triangle}).
Far from being a downside, the fact that the vertex ordering severely constrains the partial amplitudes facilitates compact combinatorial formulas for the amplitudes. 
As argued in the following section, both the distribution of the different curves in each of the fermionic trace factors and their ordering within them are completely fixed by the cubic vertex orientation.
Each cone or Feynman diagram is thus solely characterized by the set of curves that light up on the corresponding region of the Feynman fan.

\subsection{General features of the theory \label{sec:features}}

This section presents various interesting features of the partial amplitudes in the colored Yukawa theory with Lagrangian \eqref{eq:ordLagrangian}. 

\paragraph{Fermions come in adjacent pairs.}
The first property of our colored Yukawa theory we want to highlight is that all external anti-fermions are immediately followed by an external fermion in the color ordering. 
Moreover, the fermion traces always connect anti-fermion $i$ to fermion $i+1$. 
Both of the above facts are a consequence of the single vertex orientation in (\ref{eq:ordLagrangian}): if anti-fermion $i$ was instead followed by a scalar $i+1$, then the fermion line starting from $i$ would have a scalar leg emerging from the left. This vertex would have a clockwise assignment of $\bar\Psi \Phi \Psi$, which doesn't respect the ordering dictated by the Lagrangian.

\paragraph{External fermion traces.}
From the surface point of view, curves appear in a given external fermion trace if and only if one of its endpoints connects to the fermion at the end of the trace. 
Explicitly, if particles $i$ and $i+1$ are $\bar{\Psi}$ and $\Psi$ respectively, all curves $C_{i+1,j}$ must correspond to a fermionic particle. This also implies that $j$ must be an external anti-fermion, a scalar or an internal puncture in the surface. 
Meanwhile, cones involving curves that connect two fermion particles do not contribute to the amplitude. Moreover, two external punctures which are not fermions are always connected by scalar curves.

These claims follow directly from the duality between surface triangulations and Feynman diagrams. Using the first feature of our theory, we know that a $\Psi$ puncture in the surface corresponds to a region between two external adjacent fermionic legs in a Feynman diagram, which are connected by a charged line. 
\begin{center}
\includegraphics[scale=.6]{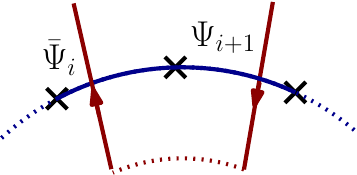}
\end{center}
Therefore, every internal propagator separating this region from any other one must be part of this line, and thus of fermionic nature. Moreover, if an internal fermion propagator separated two $\Psi$ regions, then it would be part of two different charge flows, which is obviously not possible.

\paragraph{Internal fermion traces.}
In addition to fermionic trace factors connecting two external particles, Feynman diagrams can also contain fermion lines that form closed loops. 
Such fermion trace lines are constrained to circle \emph{one and only one} internal puncture in the surface due to the orientation of the Yukawa vertex. 
Since the Feynman rules for these fermion loops involve a Dirac trace of the propagator momenta, the tropical integrand for the amplitude factorizes into separate internal traces where each curve is connected to the same internal puncture $C_{0_a,j}$. 
This is again a consequence of the vertex ordering, which forces the fermionic charge flow to stay confined to a single loop.
We already know that the puncture $j$ in $C_{0_a,j}$ cannot be an external fermion $\Psi$ (such curves are always part of external charge flows), so $j$ must be an external anti-fermion $\bar{\Psi}$, external scalar $\Phi$ or another internal puncture. 

Even so, these curves are not guaranteed to be fermionic. 
Since the Lagrangian includes a cubic scalar interaction, it is clear that for every Feynman diagram that contains a closed fermion loop we must include an analogous graph where all the propagators participating in the closed fermion loop are replaced by scalar propagators. 
In other words, each internal puncture can behave both as a fermion and a scalar. 
In the former case, all the curves connected to the puncture carry fermionic charge. 
This is easily seen with the triangulation/diagram duality, since the fermionic loop propagators separate the internal puncture region from the other ones.
This leads to the appearance of a closed Dirac trace in the corresponding Feynman diagrams. 
Moreover, the contribution of this charge assignment vanishes on the cones with curves where the internal puncture is connected to another fermion, since the two endpoints would be $\Psi$ particles. 

\begin{figure}
    \centering
    \includegraphics[width=10cm,valign=c]{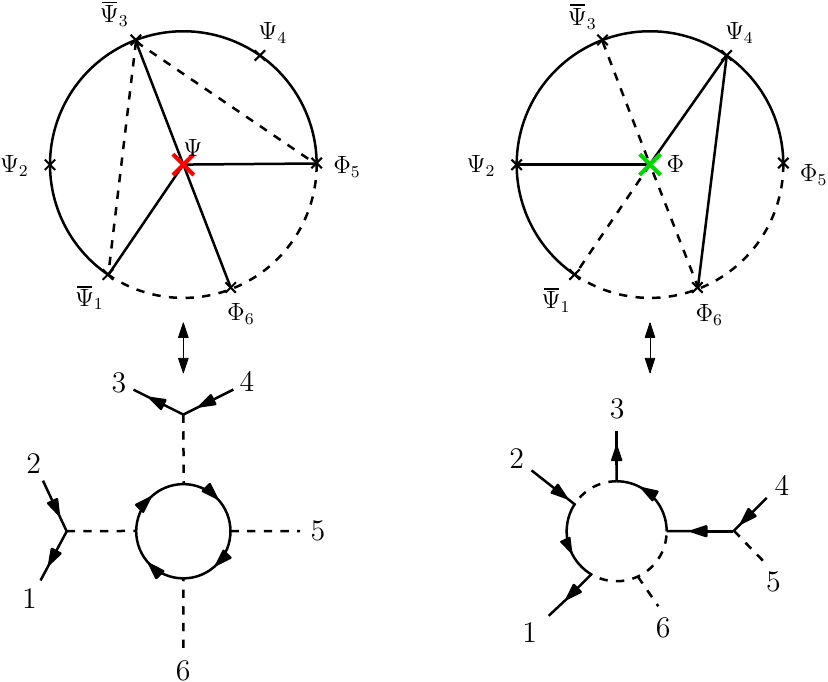} 
    \caption{On the left, the internal puncture is fixed to be fermionic. Therefore, all curves connected to it are charged and form a closed fermion line. 
    On the right, the puncture is scalar. 
    Therefore, the curves connected to it are not necessarily charged.}
    \label{fig:charge assignments} 
\end{figure}

On the other hand, when the internal puncture is not fixed to be a fermion, the curves connected to this puncture are considered fermionic if and only if they are connected to another $\Psi$ puncture.  
Otherwise, they are considered scalar. 
This leads to Feynman diagrams where the propagators in the loop are either part of a different fermion trace (for example, an open line connecting two external legs) or scalar. 
The different scenarios described above are illustrated in figure \ref{fig:charge assignments}, where the external legs of the Feynman diagrams have been assigned to punctures in the surface by moving counter-clockwise.

All in all, when writing down the numerator, we have to sum over all the possible $2^L$ charge assignments for the set of internal punctures, where $L$ is the number of closed loops in the amplitude.

Lastly, since the surface is invariant under permutations of the internal punctures, we always order the punctures such that $0_a$ comes before $0_b$ if $a<b$.

\paragraph{Universal fermion flow order.}
Interestingly, it turns out that the order in which the different curves appear inside each fermion trace is consistent across all cones in the Feynman fan, and is fully fixed by their mountainscape. 

\begin{figure}
    \centering
    \includegraphics[width=7cm,valign=c]{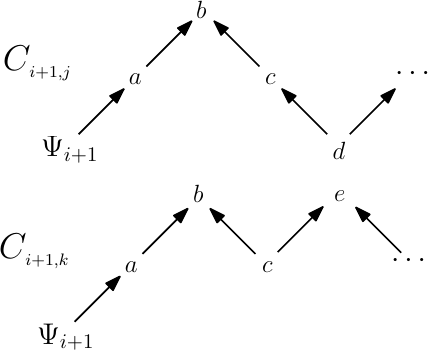} 
    \caption{The mountainscapes for two curves starting at fermion $i+1$ start differing at the turn after $c$. In the trace within the partial amplitude, the one that turns right (upper) will appear before the one that turns left (lower).}
    \label{fig:mountainscape ordering} 
\end{figure}

To see this, consider the case of an external fermion line connecting $\bar{\Psi}_i$ and $\Psi_{i+1}$, as well as two of its curves $C_{i+1,j}$ and $ C_{i+1,k}$. 
Since they have a common starting point at $i+1$, the two mountainscapes associated with the curves will have an initial part in common and differ after some specific turn (see figure \ref{fig:mountainscape ordering}). 
In particular, the momenta associated to the curve that turns right first always appears before the other one in the chain of $\slashed{P}_C$ in a spinor product $\bar v(p_i)\cdots \slashed{P}_{C_{i+1,j}} \cdots \slashed{P}_{C_{i+1,k}} \cdots u(p_{i+1})$.
Indeed, turning right first in the fatgraph means that the curve connects to a puncture in the surface that appears later in clockwise order. Using the graph duality, this means that the region corresponding to that puncture in the Feynman diagram is closer to the antifermion ($\bar\Psi$) leg, and thus will appear first in the trace factor.

For internal fermion loops, the argument is the same except that we need to choose a starting curve for the closed loop in order to compare curves. The need for such a choice arises from the cyclic symmetry of the trace (in the curve language, this reflects that curves connecting to internal punctures spiral forever around it without a definite endpoint).

\paragraph{Complete set of Feynman integrals for Yukawa.}
As discussed above, the seemingly innocent choice of only including one ordering for the cubic vertex in the Lagrangian has constrained the form of the partial amplitudes in a remarkable manner. 
If we had included the complementary direction of the charge flow, we would have to consider all the possible permutations of the curves appearing in each Feynman diagram, spoiling their distribution and ordering within the different fermion lines.

However, despite the fact that only a restricted class of Feynman diagrams contribute to a partial amplitude, 
this model still generates the full set of Feynman integrals that appear in the ``complete" ordinary Yukawa theory. 

Indeed, the independence of the interaction vertex on momenta means that any missing contribution to the partial amplitude is identical to the one corresponding to some other color ordering (see figure \ref{fig:non-planar triangle}). 
Thus, we recover the uncolored theory (up to factors of $N$) by setting all $SU(N)$ generators to identity $T^a \to \mathbbold{1}$.

In the next subsection, we use these features to write down a complete yet compact expression for the partial amplitude in colored Yukawa theory.

\subsection{The combinatorial amplitude for colored Yukawa theory \label{subsec:combinatorialamp}}

Any $n$-point, $L$-loop partial amplitude in colored Yukawa theory can be expressed as a sum over Feynman diagrams, each of them consisting of a product of propagators $p_i^2$ in the denominator, weighed by a numerator factor $\nh$ and integrated over the undetermined loop momenta $l_a$. 
The numerator $\nh$ is a product over open (closed) spinor products of the form $\bar v(p_i)\cdots \slashed{p}_j \cdots u(p_{k})$ ($\tr \big[\cdots\slashed{p}_i\cdots \big]$) coming from the internal fermionic propagators in (\ref{eq:Feynman rules}). 
Additionally, a prefactor of $ig$ or $i\lambda$ is added for each vertex in the diagram
\be \ah^{(n)} = \sum_\Gamma (ig)^{V_\Psi}(i\lambda)^{V_\Phi}\int \prod_a\frac{\d^Dl_a}{i\pi^{D/2}}\ \frac{\nh_\Gamma}{\prod_{i\in\Gamma} p_i^2}~.\ee
Meanwhile, the curve formalism of \cite{Arkani-Hamed:2023lbd,Arkani-Hamed:2023mvg} formulates the partial amplitudes of colored $\Trsu \Phi^3$ theory as an integral over a $E=(n-3+3L+2g)$-dimensional vector space called the Feynman fan
\be\label{eq:scalar integrand} \ah = \int \prod_a\frac{\d^D l_a}{i\pi^{D/2}}\int \frac{\d^E \mbf{t}}{\text{MCG}}\ e^{-S(\mbf{t})},\quad S(\mbf{t}) = \sum_C \alpha_C(\mbf{t}) P_C^2~,\ee
where $C$ are the curves that one can draw on the base fatgraph, the $\alpha_C$ are the headlight functions that select the set of curves that appear in each cone and we need to mod out by the action of the Mapping Class Group (MCG) to avoid summing over equivalent Feynman diagrams.

The extension of the curve integral to a fermionic theory like our Yukawa model requires including a numerator factor $\nh$ into the surface integrand in a way that doesn't explicitly sum over Feynman diagrams. In other words, we need to express $\nh$ as a tropical function solely referencing the set of curves $C$ one can draw in the fatgraph that characterizes the amplitude. This is surprisingly straightforward to do thanks to the stringent constraints that result from fixing the cubic vertex orientation. Firstly, we need to divide the numerator into a sum over all $2^L$ possible charge assignments (fermion or scalar) for the internal punctures in the surface. We denote each configuration by the set of punctures $\mathcal{O}= \{0_a \in \Psi\}$ ($1<a<L$) chosen to be fermionic. 
For example, at two-loops, we have the following possibilities: $\text{conf}(\Psi_0) = \{ \varnothing, \{0_1\}, \{0_2\}, \{0_1, 0_2\} \}$. 

If $(P^{(k)}_{C_{ij}})^\mu$ denotes the momentum associated to the curve connecting endpoints $i,j$ in the fatgraph (with $k$ being an abstract index to differentiate between the distinct curves that share the same endpoints), then the expression for the numerator factor $\nh$ that appears in the loop integrand is given by
\be\label{eq:tropical integrand}\ba 
\nh (\mbf{t})= (-i)^E(i\lambda)^{V-\frac{N_\Psi}{2}}(ig)^{\frac{N_\Psi}{2}}\left(\prod_{i=\bar\Psi} \PF_i(\mbf{t})\right)\left(\sum_{\mathcal{O}\in\text{conf}(\Psi_0)}\ \bar{\Theta}^\mathcal{O}_{\Psi\Psi}(\mbf{t})\prod_{a\in\mathcal{O}}\ \Ptr_a (\mbf{t})\right),\ea\ee
where 
\be\label{eq:PFermionLine}\ba
   \PF_{i\in\Psi_\text{ext}} (\mbf{t}) &= \bar v(p_i) 
    \mathcal{P}\left\{
        \underset{\substack{j= \bar\Psi,\Phi,0_\bullet}}{{\prod}^{(k)}}
        \slashed{P}_{C_{i+1,j}^{(k)}}(\mbf{t})
    \right\} 
    u(p_{i+1}) 
    \,,
\\
	 \Ptr_a (\mbf{t}) &= 	
	 - \tr\left[ \mathcal{P}\left\{\underset{\substack{j= \bar\Psi,\Phi,0_b\\ b>a}}{{\prod}^{(k)}} \slashed{P}_{C_{0_a,j}^{(k)}}(\mbf{t})  \right\} \right]
	 \,,
\ea\ee
are path ordered products of tropical functions $\slashed{P}_C(\mbf{t})$ corresponding to external fermion lines and internal closed fermion traces.\footnote{We use the following shorthand for the product over curves $$\underset{\substack{j= \bar\Psi,\Phi,0_b\\ b>a}}{{\prod}^{(k)}} = \prod_{\substack{j= \bar\Psi,\Phi,0_b\\ b>a}} \prod_k~,$$ where $(k)$ distinguishes distinct curves that share the same endpoints.}
Here, $N_\Psi$ is the number of external fermionic particles, and $V = n+2L+2g-2$ is the total number of vertices for an $n$-point amplitude with $L$ closed loops. 
Note the factor of $-1$ in the second line of \eqref{eq:PFermionLine} corresponding to the sign associated to closed fermionic loops.

The product inside the first brackets goes over all $\bar\Psi_i\, \Psi_{i+1}$ pairs that form an external fermion line, while the product inside the second brackets goes over all fermionic internal punctures $0_a\in \oh$ for each configuration. To avoid counting cones that include curves connecting two fermions (which are not allowed by the vertex ordering), we include a function
\begin{align}
    \bar{\Theta}_{\Psi\Psi}^\mathcal{O} (\mbf{t})
    = \prod_{i,j  = \Psi} (1-\Theta_{C_{ij}})
    \,,
\end{align}
where the external fermions and fermionic punctures in $\mathcal{O}$ are included in the product above. 

The tropical slashed momenta are defined as
\be \slashed{P}_C(\mbf{t}) = \begin{cases}
        -\frac{g}{\lambda}\slashed{P}_C & \text{in any cone that contains}\, C,\\
        \mathbbold{1} & \text{otherwise}.
    \end{cases}
\label{eq:tropmomenta}
\ee
Explicitly, one can write this in terms of the tropical step functions introduced in \cite{Arkani-Hamed:2024vna}
\be \slashed{P}_C(\mbf{t}) = -\frac{g}{\lambda}\slashed{P}_C \Theta_C(\mbf{t}) + \mathbbold{1}\bar\Theta_C(\mbf{t}) = \mathbbold{1} + \left(-\frac{g}{\lambda}\slashed{P}_C - \mathbbold{1}\right)\Theta_C(\mbf{t}).\ee
The factor of $g/\lambda$ is introduced at the level of the propagators to guarantee the correct coupling constant scaling on each cone in the Feynman fan. 
Masses can be incorporated into the tropical $\slashed{P}$-function 
\begin{align} \label{eq:massiveGen}
    \slashed{P}_C(\mbf{t}) 
    \to \slashed{P}_C(\mbf{t}) 
    + \Theta_C(\mbf{t}) m_C
    \,, 
\end{align}
where $m_C \to m_\Psi$ if $C$ is a fermionic curve and $m_\Phi$ otherwise. 
However, the presence of an extra term without a gamma matrix makes it hard to find a loop-integrated tropical formulation of the amplitudes. For this reason, we set all masses to zero in this work.

Both here and in the loop-integrated amplitude, it is important to orient the curve appearing in the momenta $P_C^\mu$ such that the endpoint corresponding to the $\Psi$ particle is at the beginning. This way, the momentum and charge flows have the same direction, as dictated by the Feynman rules.
Finally, the path-ordering operator $\mathcal{P}$ arranges the curves inside each trace factor according to the criteria presented in section \ref{sec:features}. 
We stress again that the ordering of the curves is a completely mechanical procedure that can be done simply by comparing the paths that each of them follows in the base fatgraph (more specifically, from their mountainscapes, see figure \ref{fig:mountainscape ordering}).

Equation (\ref{eq:tropical integrand}) provides the numerator factor that, once introduced into the loop integrand of (\ref{eq:scalar integrand}), accounts for the fermionic propagators of colored Yukawa theory. 
We can go a step further and ask if it's possible to completely integrate out the loop momenta without spoiling the curvy formulation of the amplitude. 
That is, find a tropical formula that only depends on the $t$ variables spanning the Feynman fan (and external data). 

Remarkably, once the numerator is separated according to the different charge configurations for the internal punctures, the loop-integrated expression for the amplitude takes a compact form for each charge assignment $\oh$. 
Explicitly, the result of the integral is the usual scalar exponential involving the graph Symanzik polynomials, times a numerator prefactor where the Lorentz tensor dependence is packaged into a determinant of a $|C_\Psi|\times|C_\Psi|$ matrix, where $|C_\Psi|$ is the number of fermionic curves in the assignment (external plus internal)
\begin{align}\label{eq:integrated amplitude}\ba
\ah &= 
(-i)^E(i\lambda)^{V-\frac{N_\Psi}{2}}(ig)^{\frac{N_\Psi}{2}}
\\&\quad\times 
\int\frac{\d^E\mbf{t}}{\text{MCG}}\ 
\left( 
    \prod_{i=\bar\Psi} \Pgam_i
\right)
\left( 
    \sum_{\oh \in\text{conf}(\Psi_0)}
    \bar{\Theta}_{\Psi\Psi}^\oh (\mbf{t})
    \left(
        \prod_{a \in\oh} 
         \Ptr^{(\Gamma)}_a\
    \right)
    \overline{\det\Omega}_{\mathcal{O}}
\right) 
\frac{e^{\frac{\fh_0}{\uh} {-} \zh}}{\uh^{D/2}}
    \,.
\ea\end{align}
Here, we suppress the dependence on $\mbf{t}$ in the integrand above so that it fits on one line. 
We also introduce the following shorthands for the gamma matrix versions of $\PF$ and $\Ptr$
\begin{align}\label{eq:PGamma}\ba
    \Pgam_{i \in \Psi_\text{ext}} (\mbf{t})
    &= \bar v(p_i) \mathcal{P}\left\{
        \underset{\substack{j= \bar\Psi,\Phi,0_\bullet}}{{\prod}^{(k)}}
        [\gamma^{(k)}_{i+1,j}(\mbf{t})]^{\mu_C}
    \right\} u(p_{i+1})~,
    \\
    \Ptr^{(\Gamma)}_{a} (\mbf{t})
    &= -\tr\left[ 
        \mathcal{P}\left\{
            \underset{\substack{j= \bar\Psi,\Phi,0_b\\ b>a}}{{\prod}^{(k)}}
            [\gamma_{0_a,j}^{(k)}(\mbf{t})]^{\mu_{C}}
        \right\} 
    \right]~,
\ea\end{align}
where the tropical gamma matrices are
\be 
    [\gamma_C(\mbf{t})]^{\mu_C} 
    = -\frac{g}{\lambda}
        \gamma^{\mu_C} \Theta_C(\mbf{t}) 
    + \mathbbold{1}\bar\Theta_C(\mbf{t})
    = \mathbbold{1} 
    + \left(
        -\frac{g}{\lambda} \gamma^{\mu_C} 
        - \mathbbold{1}
    \right)
    \Theta_C(\mbf{t}).
\ee
In other words, $[\gamma_C(\mbf{t})]^{\mu_C}$ is a Dirac matrix in the cones where the curve $C$ lights up, and the identity matrix in every other case.
We also define a special determinant that produces the loop-integrated tensor structures
\begin{align}
    (\overline{\det\Omega}_\oh) (\mbf{t})
    &= \left(
        \prod_{C = \Psi}\frac{\p}{\p\omega_C}
    \right)
    (\det\Omega_\oh)(\mbf{t}) \Bigg|_{\omega_C=0} 
    \,.
\end{align}
A more precise definition of $\Omega$ will be given below. 

The functions $\uh,\, \fh_0$ and $\zh$ are the usual graph Symanzik polynomials, which can be defined in terms of the external kinematics and the headlight functions $\alpha_C(\mbf{t})$ of each curve. Considering that the momentum associated to a certain curve $C$ can be decomposed as a loop-dependent and a non-loop-dependent part
\be\label{eq:curve momenta} P_C^\mu = K_C^\mu + \sum_{a=1}^L l_C^a\, l_a^\mu,\quad l_C^a\in\mathbb{Z},\ee
the Symanzik polynomials are given by
\be\label{eq:Symanzik polynomials}\ba \uh = \det\Lambda,\quad \fh_0 = \mathcal{J}^T&\tilde\Lambda \mathcal{J},\quad \zh = \sum_C K_C^2 \,\alpha_C,\\
\Lambda^{ab} = \sum_C l_C^a l_C^b \alpha_C&,\quad \mathcal{J}^{a,\mu} = \sum_C l_C^a K_C^{\mu}\alpha_C,\ea\ee
where $\tilde\Lambda = (\det\Lambda)\Lambda^{-1}$ is the adjugate matrix of $\Lambda$. Finally, the numerator structure of the amplitude is encoded in the $|C_\Psi|\times|C_\Psi|$ matrix $\Omega$, which has the following entries
\be\label{eq:tensor matrix} \Omega_{C\tilde C} = \begin{cases}
    \omega_C\left[ \bar\Theta_C(\mbf{t}) + \Theta_C(\mbf{t})\left(K_C^{\mu_C} - \sum_{C'} \left(l_{C'}^a(\Lambda^{-1})^{ab} l_C^b\right)\alpha_{C'}(\mbf{t}) K_{C'}^{\mu_C} \right) \right],& C=\tilde C,\\
    \Theta_C(\mbf{t})\Theta_{\tilde C}(\mbf{t})\left[ 1 - \frac{1}{4}\omega_C\omega_{\tilde C}\left(l_{\tilde C}^a(\Lambda^{-1})^{ab} l_C^b\right)\eta^{\mu_C\mu_{\tilde C}} \right],& C\neq \tilde C.
\end{cases}
\ee
The curves $C,\tilde C$ that enter the matrix are the ones that have a fermionic puncture (either an external leg or an internal puncture of $\oh$ for a specific configuration), oriented such that the $\Psi$ endpoint is at the beginning of the path. The $\omega_C$ are auxiliary variables associated with each fermionic curve, which are used to extract the multi-linear term of the determinant (\ref{eq:integrated amplitude}) by acting with the differential operator $\prod_C\p_{\omega_C}$. Notice that, for curves joining external punctures, $l^a_C = 0$ and the matrix elements reduce simply to $K_C^\mu = P_C^\mu$ or $\omega_C$.

In the following sections, we exemplify how to use this formula to compute partial amplitudes at several orders in the loop momenta and the 't Hooft expansion, illustrating the origin of each term in the expressions (\ref{eq:tropical integrand}) and (\ref{eq:integrated amplitude}) along the way.

\section{Examplitudes} \label{sec:examplitudes}

Having presented the generic tropical expressions for the numerator factor $\nh$ and the corresponding partial amplitude $\ah$ for the theory of colored scalars and fermions at hand, we now explicitly demonstrate its utility with several examples at both tree and loop level, in increasing order of complexity.

\subsection{Trees}\label{sec:trees}
We begin by furnishing certain tree-level examples in our colored Yukawa theory (\ref{eq:ordLagrangian}), showcasing the utility of our tropical numerator formula $\nh(\mbf{t})$ given by (\ref{eq:tropical integrand}) and highlighting the several features of the theory (as elucidated in section \ref{sec:features}) it encapsulates. 

\begin{figure}[h]
    \centering
    \includegraphics[width=0.7\textwidth]{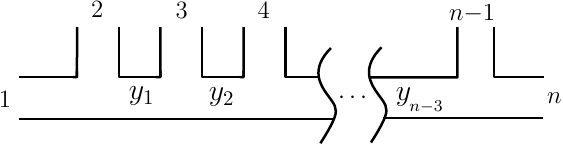}
    \caption{The $n$-point tree-level comb graph.}
    \label{fig:combgraphtree}
\end{figure}

For computing tree-level partial amplitudes, we choose the $n$-point comb graph as our distinguished fatgraph $\Gamma$, as shown in figure \ref{fig:combgraphtree}. Such a choice for $\Gamma$ with end points 1 and $n$, and with the (clockwise) ordering $123 \dots n$ allows us to list all the possible curves at tree level. For each pair of external lines $i,j$, there is a unique curve $C_{ij}$ that connects them. The paths of such curves and their associated momenta are given by 
\begin{align}
    C_{ij} &= iLy_{i-1} Ry_i Ry_{i+1} \dots Ry_{j-2}Lj  && P_{C_{ij}} = p_i + p_{i+1} + \dots + p_{j-1} \equiv p_{i, i+1, \dots, j-1}~, \nonumber \\
    C_{1j} &= 1Ry_1 Ry_2 \dots Ry_{j-2}Lj  && P_{C_{1j}} = p_{1, \dots, j-1}~, \label{eq:combgraphtreecurves} \\
    C_{in} &= iLy_{i-1} Ry_i Ry_{i+1} \dots Ry_{n-3}Rn  && P_{C_{in}} = p_{i, \dots, n-1}~, \nonumber
\end{align}
where $1<i<j-1$ and $j<n$. Using the curve matrices presented in section \ref{sec:review}, one can compute the associated headlight functions, which are given as
\begin{align}
\label{eq:treeheadlightfunc}
    \alpha_{ij}(\mbf{t}) &= \tilde f_{j-2,i} + \tilde f_{j-3,i-1} - \tilde f_{j-3,i} - \tilde f_{j-2,i-1}~, \nonumber \\ 
    \alpha_{1j}(\mbf{t}) &= -t_{j-2} + \tilde f_{j-2,1} - \tilde f_{j-3,1}~, \nonumber \\
    \alpha_{in}(\mbf{t}) &= \tilde f_{n-3,i-1} - \tilde f_{n-3,i}~,
\end{align}
with the tropical functions $\tilde f_{ij}$ defined by a reversed ordering compared to the tropical functions $f_{ij}$ given in \cite{Arkani-Hamed:2023mvg} (because our comb graph in figure~\ref{fig:combgraphtree} has the opposite orientation to their choice):
\be \tilde f_{ij}(\mbf{t}) = \text{max}(0, t_j, t_j+t_{j-1}, \dots, t_{j} + t_{j-1}+ \dots + t_i)~. \ee

Equipped with these definitions, the $n$-point tree-level partial amplitude
with $N_\Psi/2$ fermion pairs is 
\be \ah = \int \d^{n-3} \mbf{t} \ \nh(\mbf{t}) \ e^{-S(\mbf{t})}~, \label{eq:troptreeamp} \ee
where the tropical numerator formula (\ref{eq:tropical integrand}) simplifies drastically to yield
\begin{align}
\nh(\mbf{t}) = (-i)^{n-3}(i\lambda)^{n-2-\frac{N_\Psi}{2}}(ig)^{\frac{N_\Psi}{2}}  \left(\prod_{i=\bar\Psi}  \bar v(p_i)  \mathcal{P}\left\{  \prod_{j =\bar\Psi,\Phi} \slashed{P}_{i+1,j}(\mbf{t}) \right\} u(p_{i+1}) \right)~.
\label{eq:treetropintegrand}
\end{align}
The various symbols in the above expression retain their definitions as introduced in section \ref{subsec:combinatorialamp}.\footnote{We have removed the index $k$ from the tropical momenta, as every curve is uniquely determined by its endpoints at tree-level.}

\subsubsection{Five-point }

Let us now utilize the five-point example as the simplest playground to explore some of the general features of the colored Yukawa theory at action and to verify that the tropical numerator formula (\ref{eq:treetropintegrand}) explicitly reproduces the expected Feynman diagrams (cones) of the fan. 

\begin{figure}
    \centering
    \includegraphics[width=0.7\textwidth]{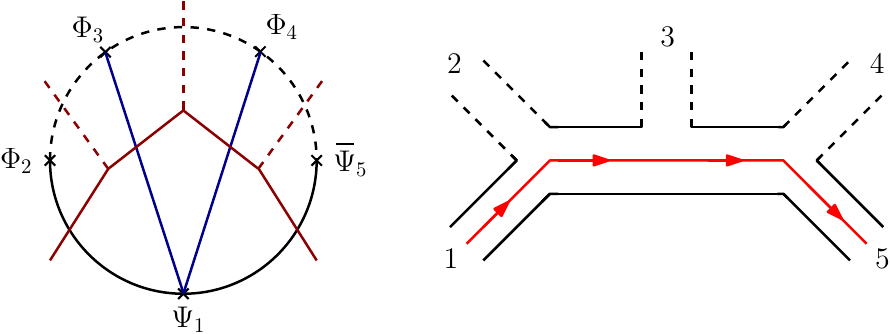}
    \caption{Triangulation ({\color{Blue}blue}) of the charged disk with fermionic (solid) internal curves leading to charge conservation at every vertex/triangle and the corresponding dual graph ({\color{Brown} red}).}
    \label{fig:correcttriangulation}
\end{figure}

Let us consider the external charge configuration $\mathcal{A}(\Psi_1,\Phi_2,\Phi_3,\Phi_4,\bar{\Psi}_5)$ involving a single fermion/anti-fermion pair and three scalars with the indicated cyclic ordering. Now, according to one of the claims presented in section \ref{sec:features}, curves $C_{13}, C_{14}$ should strictly correspond to a fermionic particle and appear in external fermion traces. Let us see why this should be the case in our example from a surface perspective. 
All curves on our base fatgraph $\Gamma$ have a fixed charge, which is determined analogously to the momentum assignment procedure:
\be Q_{C} = q_{\text{start}} + \sum_{\text{right turns}} q_{\text{incoming from the left}}~. \ee
Imposing charge conversation and the correct ordering at every vertex implies that we are left with triangles with an ordered set of charges $\{-1,0,+1\}$. Based on this observation, we see that curves $C_{13}, C_{14}$ must strictly be fermionic for this external charge assignment, as shown in the surface triangulation in figure \ref{fig:correcttriangulation}.
\begin{figure}
    \centering
    \includegraphics[width=0.7\textwidth]{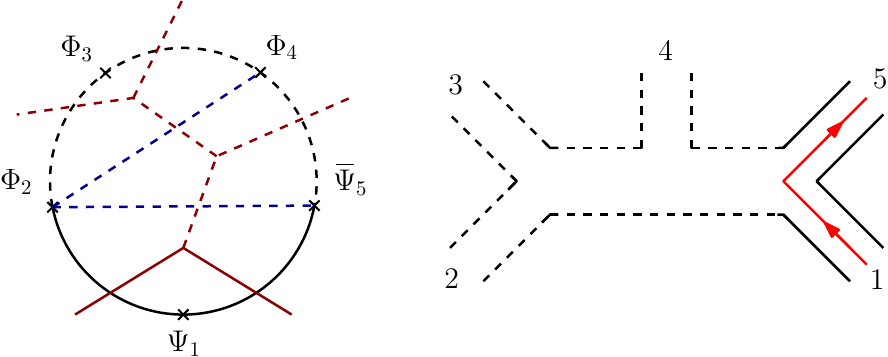}
    \caption{Triangulation ({\color{Blue}blue}) of the charged disk with scalar (dotted) internal curves leading to charge conservation at every vertex/triangle and the corresponding dual graph ({\color{Brown} red}).}
    \label{fig:allscalartriangles5pt}
\end{figure}

\begin{figure}
    \centering
    \includegraphics[width=\textwidth]{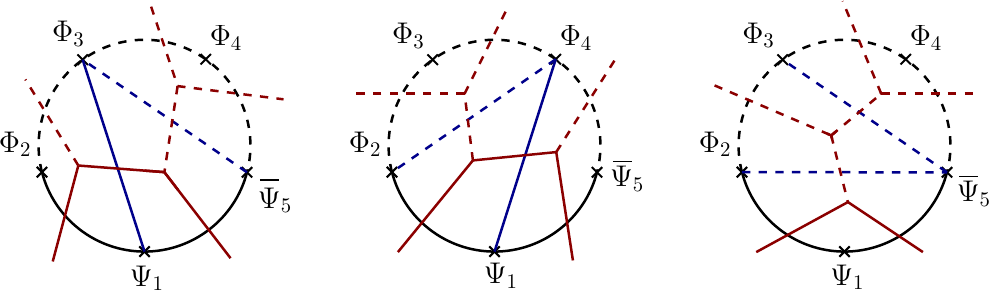}
    \caption{The remaining surface triangulations contributing to the five-point partial amplitude $\mathcal{A}(\Psi_1,\Phi_2,\Phi_3,\Phi_4,\bar{\Psi}_5)$.}
    \label{fig:remainingcones5pt}
\end{figure}

Another related claim is that two external marked points which are not fermions are always connected by scalar curves. Thus, for the arrangement of species we are considering, curves $C_{24},C_{25},C_{35}$ should be scalars. This is another consequence of imposing charge conservation at each vertex --- there will also exist triangulations containing uncharged triangles (i.e., made of only scalar (dotted) curves) corresponding to the $\Tr(\Phi^3)$ vertex in the theory, an example of which is given in figure \ref{fig:allscalartriangles5pt}. For the sake of completion,  we present the other three surface triangulations that contribute to the partial amplitude under study in figure~\ref{fig:remainingcones5pt}.

\begin{figure}
    \centering
    \includegraphics[width=0.75\textwidth]{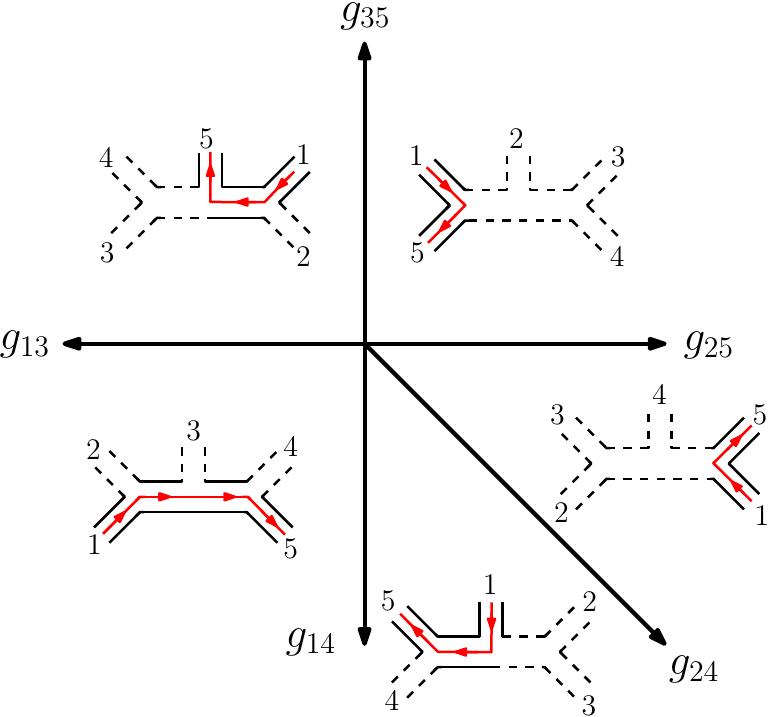}
    \caption{The tree-level Feynman fan for $\mathcal{A}(\Psi_1,\Phi_2,\Phi_3,\Phi_4,\bar{\Psi}_5)$ in colored Yukawa theory.}
    \label{fig:5ptfeynmanfan}
\end{figure}

While we have provided a bit of intuition for some of the claims presented in section \ref{sec:features} through this example, such features are inherently baked into the tropical numerator formula (\ref{eq:treetropintegrand}) --- it identifies all the fermionic/scalar curves for a given external charge assignment and churns out the associated fermion trace. For the example at hand, the expression for the tropical numerator is given by
\begin{align}
\nh(\mbf{t}) = i g \lambda^2  \left[\bar{v}(p_5) \mathcal{P}\{\slashed{P}_{C_{13}}(\mbf{t}) \slashed{P}_{C_{14}}(\mbf{t})\} u(p_1)\right] = i g \lambda^2  \left[\bar{v}(p_5) \slashed{P}_{C_{14}}(\mbf{t}) \slashed{P}_{C_{13}}(\mbf{t}) u(p_1) \right]~,
\end{align}
where we have used the universal fermion flow feature to order the curves $C_{14} > C_{13}$, which can be read off from the paths using (\ref{eq:combgraphtreecurves}). Using the relation for the tropical momenta (\ref{eq:tropmomenta}), the above expression for the tropical numerator generates the appropriate external fermionic traces corresponding to the triangulations depicted
above in figures \ref{fig:correcttriangulation}-\ref{fig:remainingcones5pt}:
\begin{align}
    \mathcal{N}\big|_{C_{13},C_{14}} &= i g^3 [\bar{v}(p_5) \slashed{p}_{123}  \slashed{p}_{12} u(p_1)] \,,
    &\mathcal{N}\big|_{C_{13},C_{35}} &= -i g^2 \lambda [\bar{v}(p_5) \slashed{p}_{12} u(p_1)] \,,
    \nonumber \\ 
    \mathcal{N}\big|_{C_{24},C_{25}} &= i g \lambda^2 [\bar{v}(p_5) \mathbbold{1} u(p_1)] \,,
    &
    \mathcal{N}\big|_{C_{14},C_{24}} &= - i g^2 \lambda [\bar{v}(p_5) \slashed{p}_{123} u(p_1)] \,,
    \nonumber \\
    \mathcal{N}\big|_{C_{25},C_{35}} &= i g \lambda^2 [\bar{v}(p_5) \mathbbold{1} u(p_1)] \,.
\end{align}
Constructing the $g$-vectors and the headlight functions $\alpha_C$ for the curves in our example using (\ref{eq:treeheadlightfunc}), we can explicitly check that our tropical expressions for the integrand and consequently the partial amplitude indeed reproduce the five Feynman diagrams in our fan, as shown in figure \ref{fig:5ptfeynmanfan}. 
Explicitly, after carrying out the integration in curve space we arrive at the following result for the partial amplitude
\begin{align}
     \mathcal{A}(\Psi_1,\Phi_2,\Phi_3,\Phi_4,\bar{\Psi}_5) &= \int \ \d^{2} \mbf{t} \ \nh(\mbf{t}) \ \exp (- \sum_C \alpha_C(\mbf{t}) P_C^2 ) \nonumber \\
     &= ig^3 \frac{[\bar{v}(p_5) \slashed{p}_{123}  \slashed{p}_{12} u(p_1)]}{p_{12}^2 p_{123}^2} - i g^2 \lambda \frac{[\bar{v}(p_5) \slashed{p}_{12} u(p_1)]}{p_{12}^2 p_{512}^2} + i g \lambda^2 \frac{[\bar{v}(p_5) u(p_1)]}{p_{23}^2 p_{234}^2} \nonumber \\
     &- i g^2 \lambda\frac{[\bar{v}(p_5) \slashed{p}_{123} u(p_1)]}{p_{23}^2 p_{123}^2} + i g \lambda^2 \frac{[\bar{v}(p_5) u(p_1)]}{p_{51}^2 p_{512}^2}~.
\end{align}
Thus, we see that our tropical formula contains all the correct contributions for this five-point tree-level partial amplitude in colored Yukawa theory as a single integral over the parameter space of the Feynman fan. 

\subsubsection{Six-point}
\begin{figure}
    \centering
    \includegraphics[width=0.7\textwidth]{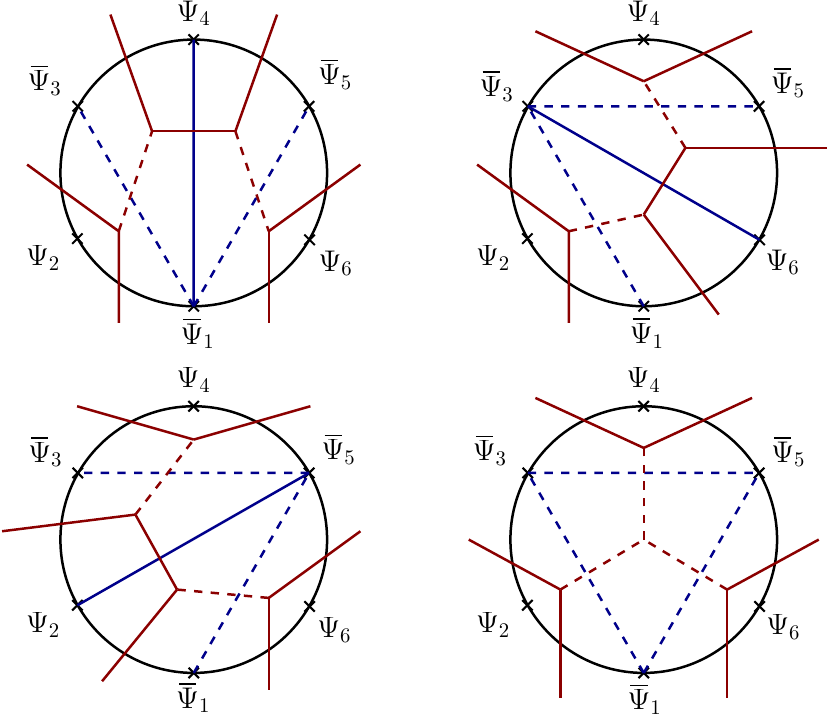}
    \caption{Surface triangulations of the charged disk with six marked points contributing to the partial amplitude $\mathcal{A}(\bar{\Psi}_1,\Psi_2,\bar{\Psi}_3,\Psi_4,\bar{\Psi}_5,\Psi_6)$.}
    \label{fig:6pttreecones}
\end{figure}
We now turn towards demonstrating a six-point tree example involving three fermionic pairs, specifically considering the partial amplitude $\mathcal{A}(\bar{\Psi}_1,\Psi_2,\bar{\Psi}_3,\Psi_4,\bar{\Psi}_5,\Psi_6)$. Due to the single ordering of the Yukawa interaction vertex, the number of diagrams/triangulations contributing to the partial amplitude is severely limited. As we shall see, there are only four contributing diagrams, a fact that is exactly reproduced by our tropical numerator formula (\ref{eq:treetropintegrand}). Moreover, following the discussion of the previous example, one should convince themselves that curves $C_{14},C_{25},C_{36}$ will be \emph{fermionic}, as they involve only a single external fermion marked point. Meanwhile, curves $C_{13},C_{15},C_{35}$ will be \emph{scalar}, as they connect external marked points which are not fermions.\footnote{The remaining curves, namely $C_{24}, C_{26}, C_{46}$, connect fermion marked points; consequently they do not contribute to the tropical formula and don't appear in valid triangulations for this amplitude.}
The tropical numerator formula for this example is given by 
\begin{align}
\mathcal{N}(\mbf{t}) = i \lambda g^3  \left[\bar{v}(p_1) \slashed{P}_{C_{25}}(\mbf{t}) u(p_2)\right] \left[\bar{v}(p_3) \slashed{P}_{C_{41}}(\mbf{t})  u(p_4)\right] \left[\bar{v}(p_5) \slashed{P}_{C_{63}}(\mbf{t})  u(p_6) \right]~. 
\end{align}
One can easily check that the above formula correctly reproduces the external trace factors appearing in the contributing cones depicted in figure \ref{fig:6pttreecones}
\begin{align}
\mathcal{N}\big|_{C_{13},C_{14},C_{15}} &= -ig^4 [\bar{v}(p_1) \mathbbold{1} u(p_2)] [\bar{v}(p_3) \slashed{p}_{456}  u(p_4)] [\bar{v}(p_5) \mathbbold{1} u(p_6)]~, \nonumber \\
\mathcal{N}\big|_{C_{13},C_{35},C_{36}} &= -ig^4 [\bar{v}(p_1) \mathbbold{1}  u(p_2)] [\bar{v}(p_3) \mathbbold{1} u(p_4)] [\bar{v}(p_5) \slashed{p}_{612} u(p_6)]~, \nonumber \\ 
\mathcal{N}\big|_{C_{15},C_{25},C_{35}} &= -ig^4 [\bar{v}(p_1) \slashed{p}_{234} u(p_2)] [\bar{v}(p_3) \mathbbold{1} u(p_4)] [\bar{v}(p_5) \mathbbold{1} u(p_6)]~, \nonumber \\
\mathcal{N}\big|_{C_{13},C_{15},C_{35}} &= ig^3 \lambda [\bar{v}(p_1) \mathbbold{1} u(p_2)] [\bar{v}(p_3) \mathbbold{1} u(p_4)] [\bar{v}(p_5) \mathbbold{1} u(p_6)]~.
\end{align}
Having worked out five-point and six-point examples explicitly to demonstrate that the tropical numerator formula generates the desired Feynman diagrams at tree level, it is clear that one can proceed in an analogous manner to work out partial amplitudes at any multiplicity.

\subsection{One-loop}\label{sec:one-loop}

In this section, we describe how the general formulas \eqref{eq:tropical integrand} and  \eqref{eq:integrated amplitude} simplify at one-loop, which we illustrate in sections \ref{sec:planar2pt} and \ref{sec:planar4pt} with the planar two- and four-point amplitudes.
Lastly, in section \ref{sec:nplanar3pt}, we compute the non-planar three-point amplitude. 

At one loop, the curve integral formula for the amplitude simplifies to 
\begin{align}
    \ah = 
    \int \d^E \mbf{t}\ \int \frac{\d^D \ell}{i\pi^{D/2}}\ \nh(\mbf{t}) e^{-S(\mbf{t})}~,
\end{align}
where, for the numerator factor, there are two charge configurations corresponding to whether the internal puncture is a scalar ($\oh = \varnothing$) or a fermion ($\oh=\{0\}$)
\be\label{eq:oneLoopIntegrand} 
\!\!\!\nh(\mbf{t}) = 
({-}i)^E 
(i\lambda)^{V{-}\frac{N_\Psi}{2}}
(ig)^{\frac{N_\Psi}{2}}
\left( 
    \prod_{i=\bar\Psi}  \PF_i
\right) 
\!\!\!
    \left( 
        \bar{\Theta}_{\Psi\Psi}^{\varnothing} 
        {-} \bar{\Theta}_{\Psi\Psi}^{0} \tr\left[ \mathcal{P}\left\{ \underset{\substack{j= \bar\Psi,\Phi}}{{\prod}^{(k)}}
        \slashed{P}_{{0,j}}^{(k)}  \right\} \right] 
    \right).\ee
Note that there is no MCG at planar one loop, and we have dropped the dependence on $\mbf{t}$ in the integrand. 
After performing the loop-momentum integration, one arrives at an expression with two contributions: one for diagrams with an internal fermion trace and one without:
\begin{align} \label{eq:oneLoopIntegrated}\ba
\ah &= 
(-i)^E 
(i\lambda)^{V-\frac{N_\Psi}{2}}
(ig)^{\frac{N_\Psi}{2}}
\int \d^E\mbf{t}\  
\left( 
    \prod_{i=\bar\Psi} \Pgam_i
\right)
\left( 
    \bar{\Theta}_{\Psi\Psi}^{\varnothing}
    \overline{\det\Omega}_{\varnothing}
    {-} \bar{\Theta}_{\Psi\Psi}^{0}
    \Ptr^{(\Gamma)}_{0}\
    \overline{\det\Omega}_{0}
\right)\
\frac{e^{\frac{\fh_0}{\uh} {-} \zh}}{\uh^{D/2}}\,.
\ea\end{align}

\subsubsection{Planar two-point \label{sec:planar2pt}}

\begin{figure}
    \centering
    \includegraphics[width=0.6\textwidth]{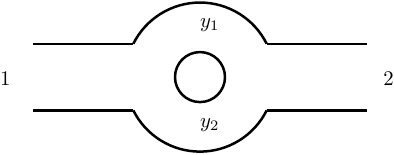}
    \caption{Base fatgraph for the two-point amplitude.}
    \label{fig:2ptBaseFatGraph}
\end{figure}

We choose the wheel topology for the base fatgraph for the planar one-loop two-point amplitude (see figure \ref{fig:2ptBaseFatGraph}). 
There are a total of four non-boundary curves on this fatgraph:
\begin{align}\ba
    C_{10} &= 1 R y_2 (L y_1 L y_2)^\infty
    \,,
    &
    C_{20} &= 2 R y_1 (L y_2 L y_1)^\infty
    \,,
    \\
    C_{11} &= 1 L y_1 R y_2 L 1
    \,,
    &
    C_{22} &= 2 L y_2 R y_1 L 2
    \,.
\ea\end{align}
The momentum associated to each curve is
\begin{align}
    P_{C_{10}} = \ell \,,
    \qquad 
    P_{C_{20}} = \ell - p_1 \,,
    \qquad 
    P_{C_{11}} = p_{12} = 0 \,,
    \qquad 
    P_{C_{22}} = p_{12} = 0 \,.
\end{align}
With this, we have all the pieces needed to use \eqref{eq:oneLoopIntegrand} for various choices of external particles.

\begin{figure}
    \centering
    \includegraphics[width=0.95\textwidth]{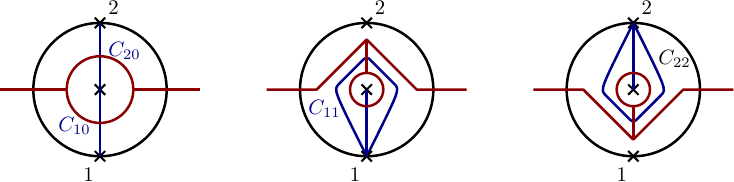}
    \caption{Triangulations of the two-point wheel surface in {\color{Blue}blue}. Each triangulation is a cone in the Feynman fan. The dual (Feynman) graph associated to each triangulation is shown in {\color{Brown}red}: a bubble and two tadpoles. The non-boundary curves have also been labeled. 
    }
    \label{fig:2ptTri}
\end{figure}

\paragraph{Example: two fermions. } 
For $1=\bar\Psi$ and $2=\Psi$, the one-loop integrand becomes
\begin{align}
    \nh(\mbf{t}) 
    = \lambda g 
    \left(
        \bar{v}(p_1) 
        \slashed{P}_{C_{20}}(\mbf{t})
        u(p_2)
    \right)
    \left(
        \bar\Theta_{22}(\mbf{t})
        - \bar\Theta_{\Psi\Psi}^{0}(\mbf{t})
            \tr\left[ \slashed{P}_{C_{01}}(\mbf{t}) \right]
    \right)\,,
\end{align}
where $\bar\Theta_{\Psi\Psi}^\varnothing = \bar\Theta_{22}$.
Note that the path ordering is not necessary in this example because one never encounters more than one $\slashed{P}_C$ in a fermion line. 
To verify that this integrand produces the expected Feynman integrals, we pull it back to the cones in the Feynman fan (see figure \ref{fig:2ptTri}).
On the Feynman cones, the integrand becomes
\begin{align} \ba
    \nh(\mbf{t}) \vert_{\text{bub}} 
    &= - g^2\
    \bar{v}(p_1) (\slashed{\ell} - \slashed{p}_1) u(p_2)
    = \includegraphics[align=c, scale=.75]{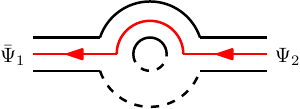}
    \,,
    \\[.5em]
    \nh(\mbf{t})\vert_{\text{tad}}
    &= g\
    \left(
        \bar{v}(p_1) 
        u(p_2)
    \right)
    \left(
        \lambda
        -g\tr\left[\slashed{\ell} \right]
    \right) 
    = \includegraphics[align=c, scale=.75]{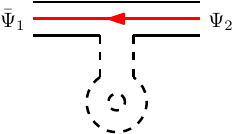} 
    + \includegraphics[align=c, scale=.75]{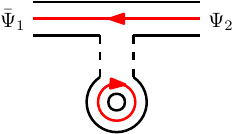}
    \,,
    \\[.5em]
    \nh(\mbf{t}) \vert_{\text{tad}^\prime}
    &= 0
    \,,
\ea\end{align}
where the above equalities hold at the level of the loop-momentum numerators. 
We have used the fact that 
$\bar\Theta_{\Psi\Psi}^{0}$ vanishes on the bubble cone while both $\bar\Theta_{\Psi\Psi}^\varnothing$ and $\bar\Theta_{\Psi\Psi}^{0}$
vanish on the last tadpole cone in figure \ref{fig:2ptTri}. 

Either working directly from the Feynman diagrams or \eqref{eq:oneLoopIntegrand}, the loop-integrated bubble with numerator evaluates to
\begin{align} \label{eq:oneLoop2ptFeyn}
    \int \frac{\d^D\ell}{i\pi^{D/2}} \int \d^E\mbf{t} \,
    \mathcal{N}(\mbf{t})\, e^{-S} \bigg\vert_{\text{bub}}
    = g^2 \bar{v}(p_1) \slashed{p}_1 u(p_2) 
    \int \d^2\alpha\,
    \frac{\alpha_1}{\alpha_1+\alpha_2} 
    \frac{
        e^{\mathcal{F}_\text{bub}/\mathcal{U}_\text{bub}}
    }{ 
        \mathcal{U}_\text{bub}^{D/2}
    }~,
\end{align}
where $\alpha_1 = \alpha_{C_{10}}\vert_\text{bub}$, $\alpha_2 = \alpha_{C_{20}}\vert_\text{bub}$, $\mathcal{U}_\text{bub} = \alpha_1 + \alpha_2$ and $\mathcal{F}_\text{bub} = - \alpha_1 \alpha_2 p_1^2$. 
Next, we illustrate how to get the above loop-integrated expression directly from the combinatorial formula \eqref{eq:oneLoopIntegrated}.

When the internal puncture is fermionic, $\oh = \{0\}$,  there is one fermionic curve: $C_{01}$ ($C_{20}$ is not allowed because it connects two fermions). 
In this case, $\Omega_0$ is the $1 \times 1$ matrix
\be\ba 
    (\Omega_0)_{C_{01}C_{01}} &= 
    \omega_{C_{01}} 
    \left[ 
        \bar\Theta_{C_{01}}(\mbf{t}) 
        + \Theta_{C_{01}}(\mbf{t}) 
            \frac{
                \alpha_{C_{20}}(\mbf{t})
                K_{C_{20}}^{\mu_{C_{01}}}
            }{\mathcal{U}}   
    \right]
    \,,
\ea\ee
Here, only $K_{C_{20}}^\mu = -p_1$ is non-vanishing ($K_{C_{01}}^\mu = K_{C_{11}}^\mu = K_{C_{22}}^\mu = 0$). 
Moreover, note that in \eqref{eq:tensor matrix}, the sum over $C^\prime$ includes \emph{all} curves, not just the fermionic ones. 
For the charge configuration $\oh = \varnothing$, the only fermionic curve is $C_{20}$ and $\Omega_\varnothing$ is also a $1\times1$ matrix 
\be\ba 
    (\Omega_\varnothing)_{C_{20}C_{20}} &= 
    \omega_{C_{20}} \left[ 
        \bar\Theta_{C_{20}}(\mbf{t}) 
        + \Theta_{C_{20}}(\mbf{t})
            \frac{
                \alpha_{C_{01}}(\mbf{t}) K_{C_{20}}^{\mu_{C_{20}}} 
            }{\mathcal{U}}
    \right]
    \,.
\ea\ee

Specifying \eqref{eq:oneLoopIntegrated} to our example, one finds that only the charge configuration $\oh = \varnothing$ contributes on the bubble cone; thus
\begin{align}
    \ah\vert_\text{bub} &=  
    -(i\lambda)(ig) 
    \int\d^E\mbf{t}\ 
    \Big( 
        \bar v(p_1) 
        [ \gamma_{C_{20}}(\mbf{t}) ]^{ \mu_{C_{20}} }
         u(p_{2})
    \Big) \
    \overline{ \det\Omega_{\varnothing } } (\mbf{t}) \
    \frac{ e^{\frac{\fh_0}{\uh} {-} \zh} }{\uh^{D/2}}
    \bigg\vert_\text{bub}~,
     \nn\\
    &=  
    g^2\ \bar v(p_1) \slashed{p}_1 u(p_{2})
    \int \d^2\alpha\ 
    \frac{\alpha_1}{\alpha_1+\alpha_2}\, 
    \frac{ e^{\fh_\text{bub}/\uh_\text{bub}} }{\uh^{D/2}_\text{bub}}\,,
\end{align}
where 
\begin{align}
    \frac{\fh_0}{\uh}-\zh\bigg\vert_\text{bub}
    =\frac{\alpha_{2}p_1^2}{\alpha_1 + \alpha_2}-\alpha_{2}p_1^2 
    =\frac{-\alpha_{1}\alpha_{2}}{\alpha_1 + \alpha_2}
    =\frac{\fh_{\text{bub}}}{\uh_\text{bub}}~,
\end{align}
and
\begin{align}
    [ \gamma_{C_{20}}(\mbf{t}) ]^{\mu_{C_{20}}}\  
    \overline{ \det\Omega^{\varnothing}} (\mbf{t}) 
    \Big\vert_\text{bub}
    = \gamma_{\mu_{C_{20}}}     
        \frac{
            \alpha_{1}
            K_{C_{20}}^{\mu_{C_{20}}}
        }{\mathcal{U}_\text{bub}}
    = \frac{
            \alpha_{1}
            \slashed{K}_{C_{20}}
        }{\mathcal{U}_\text{bub}}
    = - \frac{
            \alpha_{1}
            \slashed{p}_1
        }{\mathcal{U}_\text{bub}}
    \,.
\end{align}
Comparing to \eqref{eq:oneLoop2ptFeyn}, we find agreement.

\paragraph{Example: two scalars. } 
The other possible two-point integrand corresponds to $1=\Phi$ and $2=\Phi$:
\begin{align}
    \nh(\mbf{t}) &= 
    \lambda^2
    \left(   
            1
            -
            \tr\left[ 
                    \slashed{P}_{C_{01}}(\mbf{t})
                    \slashed{P}_{C_{02}}(\mbf{t})
            \right]
    \right)
    \,,
\end{align}
where $\Theta_{\Psi\Psi}^{\oh} = 1 \ \forall\ \mbf{t} \in \mathbb{R}^2$ since there are no external fermions.
Again, the path ordering does not matter in this example because the trace is cyclic invariant. 
Pulling back the integrand to the Feynman cones, we find the usual sum over Feynman diagrams with
\begin{align}\ba
    \nh(\mbf{t})\vert_{\text{bub}} 
    &= \includegraphics[align=c, scale=.75]{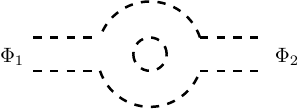}
    + \includegraphics[align=c, scale=.75]{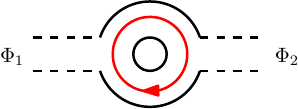}
    \,,
    \\[.5em]
    \nh(\mbf{t})\vert_{\text{tad}} 
    &= \includegraphics[align=c, scale=.75]{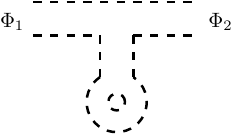}
    + \includegraphics[align=c, scale=.75]{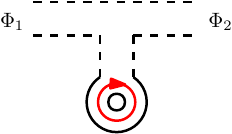}
    \,,
    \\[.5em]
    \nh(\mbf{t})\vert_{\text{tad}^\prime} 
    &= \includegraphics[align=c, scale=.75]{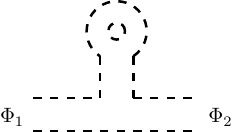}
    + \includegraphics[align=c, scale=.75]{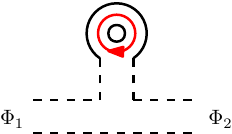}
    \,,
\ea\end{align}
where the above equalities hold at the level of the numerator of the momentum space integrand.  

Since there are no external fermions, $\Omega_\varnothing = 1$. When the puncture is fermionic there are two internal fermionic curves: $C_{01}$ and $C_{02}$. 
Therefore, 
\be\ba 
    (\Omega_0)_{C_{01}C_{01}}\vert_\text{bub} &= 
    -\omega_{C_{01}} 
    \frac{
        \alpha_{2}
        K_{C_{02}}^{\mu_{C_{01}}}
    }{\mathcal{U}_\text{bub}}   
    \,,
    \\
    (\Omega_0)_{C_{02}C_{02}}\vert_\text{bub} &= 
    \omega_{C_{02}}
    \frac{
        \alpha_1 K_{C_{02}}^{\mu_{C_{02}}} 
    }{\mathcal{U}_\text{bub}}
    \,,
    \\
    (\Omega_0)_{C_{01}C_{02}}\vert_\text{bub} &=
    1 - \frac{ \omega_{C_{01}}\omega_{C_{02}} }{4}
    \frac{\eta^{\mu_{C_{01}}\mu_{C_{02}}}}{\mathcal{U}_\text{bub}} 
    \,.
\ea\ee
On the bubble cone, the amplitude becomes 
\begin{align}\ba
    \ah\vert_\text{bub} &=  
    \lambda^2
    \int \d^E\mbf{t}\ 
    \left(
        1
        {-} \frac{g^2}{\lambda^2}
        \frac{ 
            \uh_\text{bub} 
                \tr[\gamma^\mu \gamma_\mu] 
            - 2 \alpha_1 \alpha_2 
                \tr[\slashed{p}_1 \slashed{p}_1]
        }{2 \uh_\text{bub}^2}
    \right)
    \frac{ 
        e^{ \fh_\text{bub} / \uh_\text{bub} }
    }{
        \uh_\text{bub}^{D/2}
    }
    \,.
\ea\end{align}
Here, the first term is clearly recognizable as the $\Trsu(\Phi^3)$ bubble while the second term is the bubble with a fermion trace.

\subsubsection{Planar four-point \label{sec:planar4pt}}

\begin{figure}
    \centering
    \includegraphics[width=0.45\textwidth]{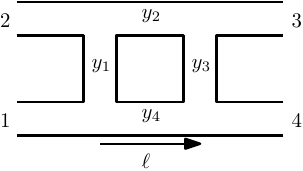}
    \caption{The one-loop four-point base fatgraph.
    }
    \label{fig:4ptWheel}
\end{figure}

\begin{figure}
    \centering
    \includegraphics[width=.8\textwidth]{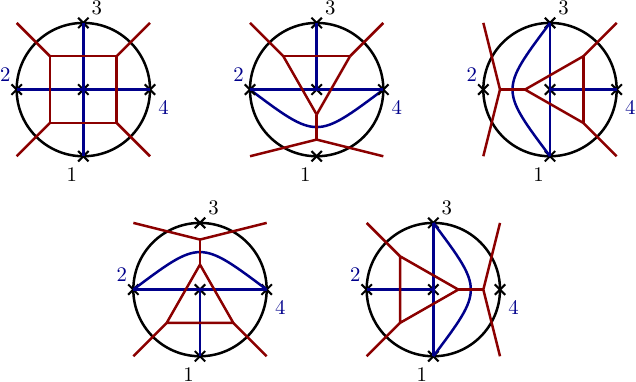}
    \caption{Some of the Feynman cones at one-loop and four points. The curves that triangulate the dual surface are depicted in {\color{Blue} blue} and the corresponding Feynman graphs in {\color{Brown} red}.
    }
    \label{fig:4ptTri}
\end{figure}

The four-point wheel fatgraph (figure \ref{fig:4ptWheel}) has a total of 35 triangulation or cones. 
Due to the large number of cones, we only present a check of our tropical formula for the box and triangle cones shown in figure \ref{fig:4ptTri}. 
Furthermore, we will only consider the amplitude with two fermions and two scalars: $\mathcal{A}(\bar{\Psi}_1,\Psi_2,\Phi_3,\Phi_4)$.

To determine the integrand numerator, only the words for the non-boundary curves connected to the puncture or the fermion $\Psi_2$ are needed:
\begin{align}\ba \label{eq:1L4ptWords}
    C_{10} &= 1 R (y_4 L y_3 L y_2 L y_1 L)^\infty
    \,,
    &
    C_{20} &= 2 R (y_1 L y_4 L y_3 L y_2 L)^\infty
    \,,
    \\
    C_{30} &= 3 R (y_2 L y_1 L y_4 L y_3 L)^\infty
    \,,
    &
    C_{40} &= 4 R (y_3 L y_2 L y_1 L y_4 L)^\infty
    \,,
    \\
    C_{24}^{\text{cc}} &= 2 R y_1 L y_4 R 4
    \,,
    &
    C_{24} &= 2 L y_2 R y_3 L y_4 
    \,,
    \\
    C_{21} &= 2 L y_2 R y_3 R y_4 L 1
    \,,
    &
    C_{23}^{\text{cc}} &= 2Ry_1Ly_4Ly_3R3
    \,.
\ea\end{align}
Here, the superscript ``$\text{cc}$" on a curve $C^{\text{cc}}_{ij}$ indicates that the curve travels from starting point $i$ to endpoint $j$ in a counterclockwise manner (i.e.~passing the puncture from the right). 
This is the first instance of the superscript $(k)$ in the formulae \eqref{eq:PFermionLine} and \eqref{eq:PGamma}. 
The momenta associated to each curve are
\begin{align}\ba
    P_{C_{10}} &= \ell
    \,,
    &
    P_{C_{20}} &= \ell-p_1
    \,,
    &
    P_{C_{30}} &= \ell-p_{12}
    \,,
    &
    P_{C_{40}} &= \ell+p_4
    \,,
    \\
    P_{C_{24}^{\text{cc}}} &= -p_{23}
    \,,
    &
    P_{C_{24}} &= p_{23}
    \,,
    &
    P_{C_{23}^{\text{cc}}} &= p_2
    \,,
    &
    P_{C_{21}} &= p_{234}
    \,.
\ea\end{align}
With this we have enough data to evaluate the tropical integrand
\begin{align}\ba \label{eq:1L4ptNum}
    \mathcal{N}(\mbf{t})
    &= g \lambda^{3}\
        \left(
            \bar v(p_1)
                \slashed{P}_{C_{24}^\text{cc}}(\mbf{t})
                \slashed{P}_{C_{23}^\text{cc}}(\mbf{t})
                \slashed{P}_{C_{20}}(\mbf{t})
                \slashed{P}_{C_{21}}(\mbf{t})
                \slashed{P}_{C_{24}}(\mbf{t})
            u(p_{2}) 
        \right)
        \\ & \qquad \times 
        \left(   
            \bar{\Theta}_{\Psi\Psi}^\varnothing (\mbf{t})
            - \bar{\Theta}_{\Psi\Psi}^0(\mbf{t})
            \tr\left[
                \slashed{P}_{C_{03}}(\mbf{t})
                \slashed{P}_{C_{04}}(\mbf{t})
                \slashed{P}_{C_{01}}(\mbf{t})
            \right]
        \right)
        \,.
\ea\end{align}
Unlike the two-point example, the path ordering is important in this case.
Examination of the words \eqref{eq:1L4ptWords} reveals that $C_{24}^\text{cc}$, $C_{23}^\text{cc}$ and $C_{20}$ turn right immediately, while $C_{24}$ and $C_{21}$ turn left first. 
Thus, $C_{24}$ and $C_{21}$ appear at the end of the fermion line.
Furthermore, since $C_{21}$ turns right before $C_{24}$ it must come first: $\bullet < C_{21} < C_{24}$. 
While $C_{24}^\text{cc}$, $C_{23}^\text{cc}$ and $C_{20}$ share the same first right turn, $C_{20}$ does not turn right again which means that it must appear after $C_{24}^\text{cc}$ and $C_{23}^\text{cc}$: $ \bullet < C_{20} < C_{21} < C_{24}$. 
Comparing the final two curves fixes the ordering of the curves in the external fermion line to be
\begin{align}
    C_{24}^\text{cc} < C_{23}^\text{cc}
    < C_{20} < C_{21} < C_{24}
    \,.
\end{align}
The ordering of the curves in the internal fermion trace is determined by the same rule but differs somewhat due to the fact there is no natural starting point for the closed fermion trace.
Picking $y_1$ as the starting road and expanding the words for the curves appearing in the fermion trace gives
\begin{align}\ba
    C_{01} = (R y_1 R y_2 R y_3 R y_4)^\infty 
    R &{\color{BrickRed} y_1 R y_2 R y_3 R y_4 L 1}
    \,,
    \\
    C_{03} = (R y_3 R y_4 R y_1 R y_2)^\infty 
    R y_3 R y_4 R &{\color{BrickRed} y_1 R y_2 L 3 }
    \,,
    \\
    C_{04} = (R y_4 R y_1 R y_2 R y_3)^\infty 
    R y_4 R &{\color{BrickRed} y_1 R y_2 R y_3 L 4}\,,
\ea\end{align}
where we see that $C_{01}$ turns right first followed by $C_{04}$ and then $C_{03}$. 
Thus, the curves in the closed fermion loop are ordered according to $C_{01} < C_{04} < C_{03}$. 

It is easy to verify that the numerator \eqref{eq:1L4ptNum} evaluates to the expected Feynman integrands. 
For illustrative purposes, we show the evaluation of this numerator on the box and second triangle cone of figure \ref{fig:4ptTri}
\begin{align}
    \mathcal{N}(\mbf{t})\vert_\text{box}
        &= -\lambda^2 g^2 \left(
            \bar v(p_1)
                \slashed{P}_{C_{20}}
            u(p_{2}) 
        \right)
        = \includegraphics[align=c,scale=.75]{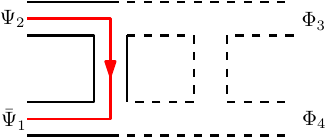}
    \,,
\end{align}
\begin{align}
    \mathcal{N}(\mbf{t})\vert_{\text{tri}_2}
        &= \lambda^3 g \ 
        \bar v(p_1) u(p_{2}) 
        \left(   
            1
            + \frac{g^3}{\lambda^3}
            \tr\left[
                \slashed{P}_{C_{01}}
                \slashed{P}_{C_{04}}
                \slashed{P}_{C_{03}}
            \right]
        \right)~,
    \nn\\[.5em] &
        = \includegraphics[align=c,scale=.75]{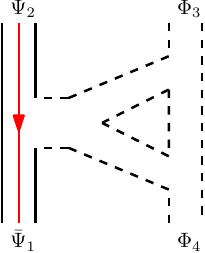}
        \quad + \quad 
        \includegraphics[align=c,scale=.75]{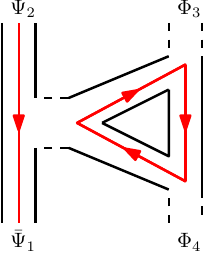}
        \,~,
\end{align}
where the above equalities are at the level of the momentum space numerator of the integrand rather than the full integral. 

Lastly, for completeness, we provide the relevant matrix elements of $\Omega$ on the box and second triangle cones:
\begin{align}
    (\Omega_\varnothing)_{C_{20} C_{20}}\vert_\text{box}&= 
    \omega_{C_{02}}\left[
        K_{C_{20}}^{\mu_{C_{20}}} 
        - \frac{
            \alpha_1 K_{C_{10}}^{\mu_{C_{10}}}  
            + \alpha_2 K_{C_{20}}^{\mu_{C_{20}}}  
            + \alpha_3 K_{C_{30}}^{\mu_{C_{30}}}  
            + \alpha_4 K_{C_{40}}^{\mu_{C_{40}}}  
        }{\uh_{\text{box}}}
    \right]
    \,,
    \\
    (\Omega_0)_{C_{0i} C_{0i}} \vert_{\text{tri}_2} &= \omega_{C_{0i}}\left[
        K_{C_{0i}}^{\mu_{C_{0i}}} 
        - \frac{
            \alpha_1 K_{C_{01}}^{\mu_{C_{0i}}} 
            + \alpha_3 K_{C_{03}}^{\mu_{C_{0i}}} 
            + \alpha_4 K_{C_{04}}^{\mu_{C_{0i}}} 
        }{\uh_{\text{tri}_2}}
    \right]
    \,, 
    \\
    (\Omega_0)_{C_{0i} C_{0j}}\vert_{\text{tri}_2} &=1 
    - \omega_{C_{0i}} \omega_{C_{0j}} 
        \frac{
            \eta^{ 
                    \mu_{C_{0i}} 
                    \mu_{C_{0j}} 
                }
        }{4}
    \,,
\end{align}
where $i,j = 1,3,4$, $\uh_\text{box} = \alpha_1 + \alpha_2 + \alpha_3 + \alpha_4$, $\uh_{\text{tri}_2} = \alpha_1 + \alpha_3 +\alpha_4$.
Other matrix elements either vanish or are simply equal to the variable $\omega_C$ in these cones.
Note that, in general, one needs to account for the fact that $\ell_{C_{0i}} = - \ell_{C_{i0}}$. 
We also remind the reader that $K_{C_{10}} = 0$, $K_{C_{20}} = -p_2$, $K_{C_{30}} = -p_{12}$ and $K_{C_{40}} = p_4$.

\subsubsection{Non-planar three-point \label{sec:nplanar3pt}}

\begin{figure}
    \centering    \includegraphics[width=.35\textwidth,valign=c]{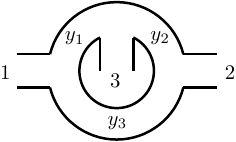} 
    \caption{Choice of base fatgraph for the non-planar three-point amplitude.}
    \label{fig:3pt non-planar base} 
\end{figure}

In this section we check how our formalism works for a non-planar example:  $\ah(\bar\Psi_1,\Psi_2\vert\Phi_3)$.
Here the vertical bar indicates that the two fermionic particles belong to a separate color trace than the scalar.
For the base fatgraph we choose the topology depicted in figure \ref{fig:3pt non-planar base}.
There are four types of non-boundary curves that can be drawn on this fatgraph: the ones connecting leg 1 to leg 3, the ones connecting leg 2 to leg 3, and the ones connecting legs 1 and 2 to themselves. From those, we need to discard the curves of the form $C_{22}$, since those would give rise to Feynman diagrams with the wrong vertex ordering (there cannot be curves with two fermions $\Psi$ as endpoints).

Moreover, it is easy to see that we can draw an infinite number of distinct curves $C_{13},C_{23}$, by circling around the loop an arbitrary number of times before exiting on leg 3. 
This leads to an overcounting of Feynman diagrams, which is a manifestation of the action of the \emph{Mapping Class Group} (MCG) \cite{Arkani-Hamed:2023lbd}. 
In order to arrive at the correct expression for the amplitude, we need to mod out by the action of this group. 
This is done with the help of the Mirzakhani kernel $\mathcal{K}$, which restricts the integration to the fundamental domain of the Feynman fan. 

In order to compute it, we start by identifying the single non-trivial closed curve that generates the MCG on the surface in figure \ref{fig:3pt non-planar base}:
\be \Delta = y_1L y_2 R y_3 R\,.\ee
Since $\Delta$ only intersects curves of the form $C_{13}$ and $C_{23}$, those are the only ones that are not invariant under the action of the corresponding Dehn twists $\gamma_\Delta$. Thus, we need to choose one representative for each coset, which we choose to be 
\be C_{13}^0 = 1L y_1 R 3\,,\quad C_{23}^0 = 2R y_2 L 3\,.\ee
Acting on these with Dehn twists $\gamma_\Delta$ generates different curves, for example
\be C_{13}^1 \equiv \gamma_\Delta(C_{13}^0)= 1 L y_1 L y_2 R y_3 R y_1 R 3,\quad C_{23}^{1}\equiv \gamma_\Delta(C_{23}^0) = 2 L y_3 R y_1 R 3.\ee 
The Mirzakhani kernel is then given by
\be \mathcal{K} = \frac{\alpha_{13}^0 + \alpha_{23}^0}{\rho}\,,\ee
where $\rho$ is a sum over all curves $C_{13},C_{23}$ that are contained in the set $\mathcal{S}$ of paths compatible with our coset representatives:
\be \rho = \sum_{C_{13}\in\mathcal{S}}\alpha_{C_{13}} + \sum_{C_{23}\in\mathcal{S}}\alpha_{C_{23}}\,.\ee
We find there are 7 curves that are compatible on the support of this Mirzakhani kernel: $C^n_{13}$, $C^n_{23}$, and $C_{11}$ with $n \in \{-1,0,+1\}$. 
Out of these, only $C^n_{23}$ are fermionic and appear inside the external fermion trace connecting particles 1 and 2 (there are no closed fermion loops, as there are no internal punctures in the surface).

\begin{figure}
    \centering
    \includegraphics[width=.9\textwidth,valign=c]{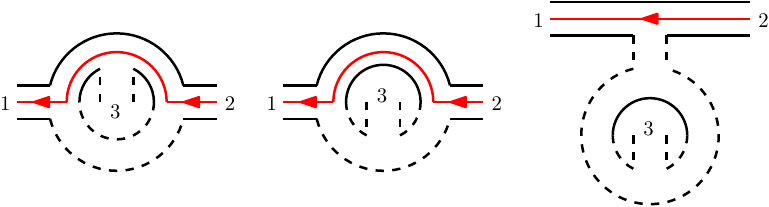} 
    \caption{Feynman diagrams appearing for the non-planar three-point amplitude.}
    \label{fig:non-planar 3pt topologies} 
\end{figure}

Explicitly, the tropical numerator becomes
\be \nh = \lambda^2 g\,\bar\Theta_{C_{22}}(\mbf{t})\bar v(p_1) \slashed{P}_{C_{23}^{-1}}(\mbf{t})\slashed{P}_{C_{23}^0}(\mbf{t})\slashed{P}_{C_{23}^1}(\mbf{t}) u(p_2)\,.\ee
We can go one step further and integrate out the loop momentum. Since there are no potential closed fermion loops, there is only one possible tropical tensor structure in the numerator, which is given by the $3\times 3$ matrix with entries described in (\ref{eq:tensor matrix}):
\begin{align}
(\Omega_\varnothing)_{nn} &= \omega_n\Bigg[ \bar\Theta_{C_{23}^n}(\mbf{t}) {+} \Theta_{C_{23}^n}(\mbf{t})\Bigg( K_{C_{23^n}}^{\mu_n} {-} \frac{1}{\uh}\Big( \sum_{C_{13}\in\mathcal{S}}\alpha_{C_{13}}K_{C_{13}}^{\mu_n} {+} \sum_{C_{23}\in\mathcal{S}}\alpha_{C_{23}}K_{C_{23}}^{\mu_n} \Big) \Bigg) \Bigg]
\,, 
\nn\\ 
(\Omega_\varnothing)_{nm} &= \Theta_{C_{23}^n}(\mbf{t})\Theta_{C_{23}^m}(\mbf{t})\Bigg(1 - \frac{\omega_n\omega_m}{4\uh}\eta^{\mu_n\mu_m}\Bigg)
\,,
\end{align}
where $ m \neq n$. 
Above, we have condensed the notation by using the indices $n,m$ to indicate the instance of the curve $C_{23}^n$, since those are the only ones that appear in the open fermion line. As we saw in the other one-loop examples, the first Symanzik polynomial takes the simple form
\be \uh = \sum_{C_{13}\in\mathcal{S}}\alpha_{C_{13}} {+} \sum_{C_{23}\in\mathcal{S}}\alpha_{C_{23}}\,.\ee
Expanding the determinant of the matrix $\Omega_\varnothing$ and taking the coefficient of the monomial $\omega_{-1}\omega_{0}\omega_1$ yields all the possible tensor contractions after loop integration. 
Integrating over the Feynman fan by substituting the explicit expressions for the headlight and step functions $\alpha_C,\Theta_C$, we find the sum over all possible Feynman diagrams: the three different topologies are depicted in figure \ref{fig:non-planar 3pt topologies}.

%

\subsection{Planar two-loop two-point \label{sec:2loop}}

In this section we present our final example: the two-loop two-point planar amplitude.

\begin{figure}
    \centering
    \includegraphics[width=.55\textwidth,valign=c]{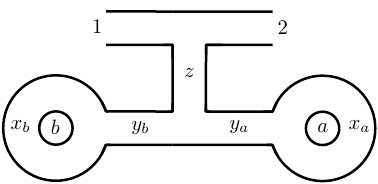} 
    \caption{Choice of base fatgraph for the planar two-loop two-point amplitude.
    This fatgraph is obtained by gluing one of the legs of a three-point tree with a two-loop tadpole fatgraph and is particularly convenient for enumerating the curves.
    }
    \label{fig:2loop2pt fatgraph} 
\end{figure}

We start by choosing the base fatgraph of figure \ref{fig:2loop2pt fatgraph}. 
This choice is especially useful since one can recycle previous quantities via tree-loop factorization \cite{Arkani-Hamed:2023mvg}. 
For example, all of the necessary surfaceology calculations (headlight functions, Symanzik polynomials, etc.)
for higher point two-loop surfaces share the same loop contributions from the tadpole subdiagrams. 
It is also easy to adapt the tree-level part by adding more external legs to the comb fatgraph of section \ref{sec:trees}.

For this surface there are three main types of curves: the ones that start at an external leg and end in a spiral around one of the loops, the ones that start and end at an external leg, and the ones that start and end spiraling. Taking into account that particle 2 is a fermion $\Psi$, we ignore every curve that connects leg 2 to itself since they lead to Feynman diagrams with the wrong vertex ordering.  

Like in section \ref{sec:nplanar3pt}, we also need to quotient out by the action of the mapping class group for surfaces with two punctures. 
Indeed, the tadpole subdiagram can be traversed by one non-trivial closed curve $\Delta$ that loops around both internal punctures
\be \Delta = y_b R x_b R y_b R y_a R x_a R y_a L\,.\ee
Since $\Delta$ intersects the curves that start at an external leg and end spiraling, this subset transforms non-trivially under the action of the MCG. 
All other curves are invariant under the MCG action since these curves do not intersect $\Delta$.

In other words, there are four MCG cosets, represented by curves of the form $C_{10_a}$, $C_{20_a}$, $C_{10_b}$ and $C_{20_b}$. 
To mod out by the action of the group, we need to identify one representative curve for each coset. We choose the shortest paths ending on a counterclockwise spiral
\be\ba C_{10_a}^0 = 1 R z L y_a R (x_a L)^\infty\,, \quad C_{20_a}^0 = 2 L z L y_a R (x_a L)^\infty\,,\\ C_{10_b}^0 = 1 R z R y_b R (x_b L)^\infty\,, \quad C_{20_b}^0 = 2 L z R y_b R (x_b L)^\infty\,.\ea\ee
Acting with Dehn twists $\gamma_\Delta$ on these curves
generates all curves not invariant under the MCG.
For example
\be C_{10_a}^1 \equiv \gamma_\Delta(C_{10_a}) = 1 R z R y_b R x_b R y_b R y_a R (x_a L)^\infty\,.\ee
The Mirzakhani kernel is then the tropical function that quotients out the MCG by selecting only the curves compatible with our representatives:
\be \mathcal{K} = \frac{\alpha_{10_a}^0+\alpha_{20_a}^0+\alpha_{10_b}^0+\alpha_{20_b}^0}{\rho}.\ee
Here, $\rho$ is the sum over all curves that intersect $\Delta$ and are compatible with $C^0_{1a}$, $C^0_{2a}$, $C^0_{1b}$ and $C^0_{2b}$ 
(as before, we call the set of all such curves $\mathcal{S}$):
\be \rho = \sum_{C_{10_a}\in\mathcal{S}}\alpha_{C_{10_a}}+\sum_{C_{20_a}\in\mathcal{S}}\alpha_{C_{20_a}}+\sum_{C_{10_b}\in\mathcal{S}}\alpha_{C_{10_b}}+\sum_{C_{20_b}\in\mathcal{S}}\alpha_{C_{20_b}}.\ee
In total, $\mathcal{S}$ contains 24 curves compatible with the Mirzakhani kernel that respect the Yukawa vertex ordering. 

Since this amplitude is planar and the internal punctures in a surface don't carry any momentum, the kinematic assignments of the curves take a simple form, and are invariant under the action of the MCG. In particular, we choose the loop parametrization
\be P^\mu_{C_{10_a}} = l_a\,,\ P^\mu_{C_{20_a}} = l_a - p\,, \ P^\mu_{C_{10_b}} = l_b\,,\ P^\mu_{C_{20_b}} = l_b - p\,,\ee
where we are referring to arbitrary curves with the indicated endpoints.

For the loop-integrated form of the tropical numerator, we make use of the matrix $\Lambda$ that enters in the expression for the Symanzik polynomials \eqref{eq:Symanzik polynomials}. To determine it, we use the telescopic property of the headlight functions \cite{Arkani-Hamed:2023mvg}, according to which the expression for the $\alpha$'s of the tadpole subgraph is simply given by the sum over all possible ways to extend them into the tree diagram. 
Therefore, we find
\be \Lambda = \begin{pmatrix}
    \alpha_{0_a0_b}+\alpha_{10_a}+\alpha_{20_a}& -\alpha_{0_a0_b}\\ -\alpha_{0_a0_b}& \alpha_{0_a0_b}+\alpha_{10_b}+\alpha_{20_b}
\end{pmatrix},\ee
where
\be \alpha_{i0_a} = \sum_{C_{i0_a}\in\mathcal{S}}\alpha_{C_{i0_a}},\ \alpha_{i0_b} = \sum_{C_{i0_b}\in\mathcal{S}}\alpha_{C_{i0_b}},\quad i=1,2\,.\ee 
As seen in previous examples, one advantage of our formulation for the Yukawa numerator is that all the tropical objects, such as the headlight functions, do not depend on the species for the external particles that characterize a specific amplitude. Thus, we can reuse these results for amplitudes containing different particle species.

Let's start with the case where the two external legs are fermionic, $\ah(\bar\Psi_1,\Psi_2)$. The charge assignments for each curve are also (partially) fixed thanks to the constraints from our Yukawa vertex ordering. 
For instance, curves of the form $C_{2j}$ ($j=1,a,b$) are always fermionic and only appear as part of the external fermion line $\bar{v}(p_1)\cdots u(p_2)$,
while curves connecting leg 1 to itself are scalar. On the other hand, the species of the curves that end on a spiral and start at either leg 1 or the other spiral depends on the specific charge configuration that is assigned to the internal punctures. 
For example, curves that begin/end at a spiral around a fermionic puncture always appear inside internal Dirac traces $\tr[\cdots\gamma\cdots]$ as part of the closed fermion loop.

\begin{figure}
    \centering
    \includegraphics[width=.4\textwidth,valign=c]{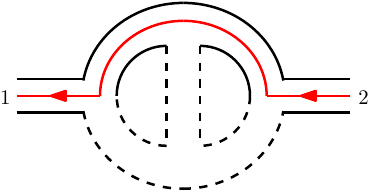} 
    \caption{Kite topology for the planar two-loop two-point fermion amplitude.}
    \label{fig:double triangle fermion} 
\end{figure}

With this, we can insert the tropical functions into the general formula (\ref{eq:tropical integrand}) to obtain an expression for the curve integral numerator of the two-loop two-point amplitude. 
Localizing to the different cones in the Feynman fan, we find the expected Feynman diagrams (there are a total of 52) of our colored Yukawa theory. 
For example, the cone corresponding to the kite topology (see figure \ref{fig:double triangle fermion}) evaluates to 
\be\label{eq:double triangle numerator} \nh\big\vert_\text{kite}= -i g^3\lambda\bar v(p_1) \slashed{P}_{C_{20_b}^0}\slashed{P}_{C_{20_a}^0} u(p_2) + \text{MCG perm.}, \ee
where the rest of the terms arise from considering spiral curves related by the MCG.
Each of these MCG-equivalent curves evaluates to the same expression once the physical value for the curve momenta $P_C$ is substituted.
This overcounting is compensated by the Mirzakhani kernel.

Let's also explicitly write the loop-integrated expression for this Feynman diagram by evaluating our formula (\ref{eq:integrated amplitude}) on the respective cones. 
By inserting the expression for $\Lambda$ into the matrix (\ref{eq:tensor matrix}) and evaluating on particular cones, we obtain the tensor structure that arises after integrating out the loop momenta in terms of the curve integral variables. However, it is also important to note that specific Feynman diagrams are associated to only one of the $2^L$ (in this case, four) possible charge assignments for the internal punctures. 

For the kite diagram, both punctures $a$ and $b$ are scalar ($\oh=\varnothing$), which is reflected by the fact that there are no closed fermion lines.
Localizing to its cone, we see that the curves that don't appear reduce to trivial factors of $\omega_C$ on the diagonal of the tensor matrix, while the non-trivial minor is given by
\be\ba 
    (\Omega_\varnothing)_{aa}
    \to& \omega_{20_a}\Bigg\{K_{C_{20_a}}^{\mu_a}-\frac{1}{\uh}\Bigg[\big(\alpha_{10_b}^0+\alpha_{20_b}^0+\alpha_{0_a0_b}\big)\big(\alpha_{10_a}^0 K_{C_{10_a}}^{\mu_a}+\alpha_{20_a}^0 K_{C_{20_a}}^{\mu_a}\big)\\ &\qquad- \big(\alpha_{10_b}^0+\alpha_{20_b}^0\big)\alpha_{0_a0_b} K_{C_{0_a0_b}}^{\mu_a} + \alpha_{0_a0_b}\big(\alpha_{10_b}^0 K_{C_{10_b}}^{\mu_a}+\alpha_{20_b}^0 K_{C_{20_b}}^{\mu_a}\big)\Bigg]\Bigg\}\,, 
    \\ 
    (\Omega_\varnothing)}_{bb
    \to& 
        (\Omega_\varnothing)_{aa}
        \Big|_{a\leftrightarrow b}\,,
    \qquad
    (\Omega_\varnothing)_{ab}
    =
    (\Omega_\varnothing)_{ba}
    \to 1-\frac{1}{4}\omega_{20_a}\omega_{20_b}\frac{\alpha_{0_a0_b}}{\uh}\eta^{\mu_a\mu_b}\,,
\ea\ee
where we have used the shorthand notation for the indices $a,b$ to correspond to the two curves $C_{20_a}^0, C_{20_b}^0$ associated with fermionic propagators in (\ref{eq:double triangle numerator}). Computing the determinant of this matrix and extracting the multilinear coefficient in $\omega_{20_a}\omega_{20_b}$, we obtain all the contributions to the tensorial dependence of the numerator after integrating the loop momenta.

Next we treat the two-loop two-point scalar amplitude $\ah(\Phi_1,\Phi_2)$. 
As remarked earlier, half of the work is already done since the expressions for the headlight functions and surface Symanzik polynomials are the same for both  $\ah(\bar\Psi_1,\Psi_2)$ and $\ah(\Phi_1,\Phi_2)$.
However, the possible charge assignments for the curves change. 
The curves of the form $C_{12}, C_{11}, C_{22}$ are all scalar, while the species of any curve that ends at a spiral is dictated by the charge assignments of the internal punctures. 
In particular, this means that fermion lines are always closed, appearing as Dirac traces of propagators involving a single loop.

\begin{figure}
    \centering
    \includegraphics[width=.4\textwidth,valign=c]{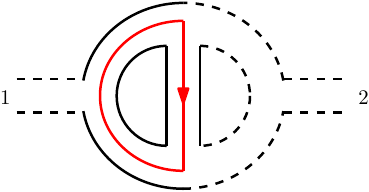} 
    \caption{Kite topology for the planar two-loop two-point scalar amplitude.}
    \label{fig:double triangle scalar} 
\end{figure}

For example, the tropical numerator evaluates to different versions of the kite diagram depending on the charge of the punctures. 
In the case that the punctures $a$ and $b$ are chosen to be fermionic and scalar, respectively ($\oh = \{0_a\}$), we find the diagram in figure \ref{fig:double triangle scalar} on the kite cone. 
On this cone, the expression for the tropical numerator becomes
\be\label{eq:double triangle scalar numerator} \nh\big\vert_\text{kite} = -i g^3\lambda\tr\big[ \slashed{P}_{C_{10_a}}\slashed{P}_{C_{0_a0_b}}\slashed{P}_{C_{20_a}} \big]\,.\ee
After loop integration, the tensor matrix for this charge assignment has a non-trivial $3\times3$ minor (again, we will abuse notation and specify the indices of the matrix in terms of the other endpoint of the curves in (\ref{eq:double triangle scalar numerator}) that end spiraling around $a$) with entries
\begin{align}
    &(\Omega_{0_a})_{11}
    \to \omega_{10_a}\Bigg\{K_{C_{0_a1}}^{\mu_1}-\frac{1}{\uh}
    \Bigg[
        \big(
            \alpha_{10_b}^0
            {+} \alpha_{20_b}^0
            {+} \alpha_{0_a0_b}
        \big)
        \big(
            \alpha_{10_a}^0 
                K_{C_{0_a1}}^{\mu_1}
            {+} \alpha_{20_a}^0 
                K_{C_{0_a2}}^{\mu_1}
        \big)
    \nn\\ &\qquad\qquad\qquad {+}
    \big(
        \alpha_{10_b}^0
        {+} \alpha_{20_b}^0
    \big)\alpha_{0_a0_b} K_{C_{0_a0_b}}^{\mu_1} 
    {+} \alpha_{0_a0_b}
    \big(
        \alpha_{10_b}^0 K_{C_{0_b1}}^{\mu_1}
        {+}\alpha_{20_b}^0 K_{C_{0_b2}}^{\mu_1}
    \big)
        \Bigg]
    \Bigg\}\,, 
\nn\\ 
    &(\Omega_{0_a})_{bb}
    \to \omega_{0_a0_b}\Bigg\{K_{C_{0_a0_b}}^{\mu_b}{-}\frac{1}{\uh}\Bigg[\big(\alpha_{10_b}^0{+}\alpha_{20_b}^0\big)\big(\alpha_{10_a}^0 K_{C_{0_a1}}^{\mu_b}{+}\alpha_{20_a}^0 K_{C_{0_a2}}^{\mu_b}\big)
    \nn\\& \qquad\qquad {+} \big(\alpha_{10_a}^0{+}\alpha_{20_a}^0{+}\alpha_{10_b}^0{+}\alpha_{20_b}^0\big)\alpha_{0_a0_b} K_{C_{0_a0_b}}^{\mu_b} 
    {-} \big(\alpha_{10_a}^0{+}\alpha_{20_a}^0\big)\big(\alpha_{10_b}^0 K_{C_{0_b1}}^{\mu_b}{+}\alpha_{20_b}^0 K_{C_{0_b2}}^{\mu_b}\big)\Bigg]\Bigg\}\,, 
\nn\\ 
    & (\Omega_{0_a})_{12}
    = (\Omega_{0_a})_{21}
    \to 1 - \frac{1}{4}\omega_{10_a}\omega_{20_a}\frac{\alpha_{10_b}+\alpha_{20_b}+\alpha_{0_a0_b}}{\uh}\eta^{\mu_1\mu_2}
    \,,
\nn\\ 
    & (\Omega_{0_a})_{1b} = (\Omega_{0_a})_{b1}
    \to 1+ \frac{1}{4}\omega_{10_a}\omega_{20_a}\frac{\alpha_{10_b}+\alpha_{20_b}}{\uh}\eta^{\mu_1\mu_b},
\nn\\
    & (\Omega_{0_a})_{22}
    \to \left. 
        (\Omega_{0_a})_{11}
    \right|_{1\leftrightarrow2}\,,
    \qquad
    (\Omega_{0_a})_{2b}=(\Omega_{0_a})_{b2}
    \to \left. 
        (\Omega_{0_a})_{1b}
    \right|_{1\leftrightarrow2}
    \,.
\end{align}
As in the fermion case, expanding the determinant and picking the multilinear term in $\omega_{10_a}\omega_{20_a}\omega_{0_a0_b}$ yields all the possible pairwise tensor contractions that can appear after integrating the loop momenta.

\section{Outlook}\label{sec:Outlook}

This paper extends the curve integral formalism introduced in recent works \cite{Arkani-Hamed:2023lbd,Arkani-Hamed:2023mvg,Arkani-Hamed:2024vna} to quantum field theories of colored interacting fermions and scalars. 
By making use of constraints apparent from the surface point of view, we find a compact expression for colored Yukawa amplitudes as a single integral over a tropical numerator that makes no reference to Feynman diagrams. Moreover, we can perform the loop integration and obtain expressions in terms of special determinants. 
In addition to the explicit checks of equations \eqref{eq:tropical integrand} and \eqref{eq:integrated amplitude} included in section \ref{sec:examplitudes}, we have also performed various checks of our formula for (7-10)-point tree amplitudes, planar 5- and 6-point one-loop amplitudes, non-planar 4-point amplitudes and two-loop 4-point amplitudes.

While our work provides further evidence that a combinatorial formulation of (close to) real-world scattering amplitudes is tangible, there are still numerous aspects of the curve integral framework that need to be understood better. For example, it is essential to account for massive fermions without breaking the nice combinatorial structure observed in \eqref{eq:integrated amplitude} in order to extend the curve integral formalism to more realistic theories. 
While the determinant in the loop-integrated formula does not seem robust enough to capture the generalization to non-zero fermion mass, 
its simple realization at the level of the loop integrand suggests that a combinatorial formulation is feasible. 
Moreover, to describe quantum electrodynamics or quantum chromodynamics, a curve integral formulation for amplitudes involving gauge bosons (photons or gluons) coupled to fermions in the fundamental representation of $SU(N)$ is necessary. 

Another intriguing question is whether there exists a Feynman fan more naturally suited to theories with fermions that does not require us to sum over different charge configurations for the internal punctures. In other words, a division of the tropical space that spans the complete set of Yukawa Feynman diagrams in a non-overlapping way.
This would effectively reduce the $2^L$-term sum into a tropical product.

Finally, it would be interesting to find a \emph{stringy} version of Yukawa amplitudes that reduces to equations \eqref{eq:tropical integrand} and \eqref{eq:integrated amplitude} in the low-energy limit $\alpha'\ll 1$ along the lines of \cite{Arkani-Hamed:2023swr,Arkani-Hamed:2019mrd}. 
With such a formula one could more easily test factorization in fermionic theories as well as investigate the possibility of kinematic shifts analogous to the $\delta$-shifts that relate NLSM and Yang-Mills amplitudes to their $\Trsu(\Phi^3)$ counterparts \cite{Arkani-Hamed:2023jry,Arkani-Hamed:2024nhp,Arkani-Hamed:2024fyd}.


\section*{Acknowledgements}

We are grateful to N.~Arkani-Hamed, C.~Figueiredo, H.~Frost, and G.~Salvatori for carefully explaining many aspects of their work and to S.~Paranjape and H.~Thomas for related discussions. This work was supported in part by the US Department of Energy under contract DE-SC0010010 Task F (SD, MSp, AV), by Simons Investigator Award \#376208 (AP, AV), and by Bershadsky Distinguished Visiting Fellowships at Harvard (MSp, AV).

\bibliographystyle{JHEP}
\bibliography{bibliography}
\end{document}